%% file: main_new.tex
\documentclass[aps,twocolumn,prx,superscriptaddress]{revtex4-2}
\usepackage[utf8]{inputenc}
\usepackage{graphicx}
\usepackage{tabularx}
\usepackage{booktabs}
\usepackage{multirow}
\usepackage{comment}
\usepackage{amsmath}
\usepackage{amsfonts}
\usepackage{amssymb}
\usepackage{bm}%for bold vector letters

\usepackage{tikz}
\usepackage{xcolor}

\usepackage{mymacros}
\renewcommand{\selectlanguage}[1]{}%removes error regarding language field in bibtex file
\renewcommand{\paragraph}[1]{\textit{\textbf{#1}}}
\usepackage[colorlinks=true,citecolor=blue,linkcolor=blue,urlcolor=blue]{hyperref}%% \hypersetup{
\begin{document}

\title{Control of morphology and topology in a lattice model of branching morphogenesis}
\author{Christian Hanauer}
\affiliation{Max Planck Institute for the Physics of Complex Systems, Dresden, Germany}

\author{Frank Jülicher}
\email{julicher@pks.mpg.de}
\affiliation{Max Planck Institute for the Physics of Complex Systems, Dresden, Germany}
\affiliation{Center for Systems Biology, Dresden, Germany}
\affiliation{Cluster of Excellence Physics of Life, Technische Universität Dresden, Dresden, Germany}

\author{Efe Ilker}
\email{efe.ilker@inserm.fr}
\affiliation{Max Planck Institute for the Physics of Complex Systems, Dresden, Germany}
\affiliation{Aix-Marseille Université, INSERM, DyNaMo, Turing Centre for Living Systems, Marseille 13009, France}

\date{\today}
\begin{abstract}	
We present a lattice model for morphogen-controlled branching morphogenesis which combines ideas and concepts from non-equilibrium physics and developmental biology. In this model, the stochastic occupation dynamics of cells is coupled with signaling molecules (morphogens) produced by the cells. We investigate growth patterns governed by morphogen concentration gradients, spanning regimes ranging from the diffusion-limited aggregation limit to the stochastic surface growth (Eden model) limit. Moreover, we introduce control over topology by a local operator and study growth, degrowth, and steady-state dynamics of branched patterns. The topology-preserving steady-state clusters exhibit a power-law scaling of radius of gyration with cluster size, yielding the exponents $0.68\pm0.01$ in square lattice and $0.67\pm 0.01$ in hexagonal lattice. 
\end{abstract}
\maketitle

Branched patterns appear in many biological systems including vasculature \cite{lorthois_2010,family_1989,mainster_1990}, the lung \cite{weibel_1963,weibel_1991}, and neurons \cite{shree_2022}. For various branched systems in biology, the geometry and topology are relevant for function. Tree-like networks can provide optimized transport properties \cite{west_1997,banavar_2000}, while the presence of loops can introduce redundancies that can lead to resilience with respect to damage and facilitate accommodating load fluctuations \cite{katifori_2010,corson_2010}. 
Branched biological networks are typically multicellular systems which grow by
cell growth and division. Interestingly, some systems such as flatworm can also 
reverse growth and undergo degrowth when starved and cells die \cite{rink2013stem}.
Open problems include the questions of how
network topology is controlled, and how
growth and degrowth of branched patterns can be controlled, and how steady branched patterns can be maintained  \cite{le_verge_2024}.

The formation of branched structures is a well-studied problem in non-equilibrium physics \cite{langer_1980,vicsek_1992,meakin_1998}. A classic model for the growth of fractal, branched patterns is diffusion-limited aggregation (DLA) \cite{witten_1981,witten_1983}. In DLA, the growth process starts with a seed particle in the center of a lattice. Next, a random walker is released far away from the seed and moves on the lattice until it arrives at a site next to the seed where it is irreversibly attached. Repeating this process leads to the growth of a cluster forming branched structures that are self-similar and can be characterized by fractal dimensions and scaling behaviors. This simple model can account for key features of pattern formation in a variety of systems ranging from dielectric breakdown \cite{niemeyer_1984}, viscous fingering \cite{paterson_1984,maloy_1985}, electrodeposition \cite{matsushita1984fractal}, and the growth of bacterial colonies \cite{fujikawa1989fractal,matsushita1990diffusion,fujikawa1991bacterial}. Work in recent decades has revealed effects of anisotropic growth rules, lattice anisotropies, cross-over phenomena affecting the self similarity for large lattices \cite{kesten1987long,meakin1995progress,halsey2000diffusion,grebenkov2017anisotropy}, as well as contrasts with off-lattice formulations \cite{tolman1989off,hastings1997renormalization,hastings1998laplacian}. 

A different class of branched structures are branched polymers. They can be studied in lattice models where they are called lattice animals. \cite{lubensky1979statistics,parisi1981critical,you1998critical,jensen2000statistics,hsu2005simulations}. These branched structures can have different topology than DLA clusters and give rise to different scaling when studied in good solvent conditions.  

In contrast to such physical branching processes, branched structures in biological systems differ in several respects. 
Biological branched patterns show self-similarity and scaling but are not perfect fractals.
Branching in multicellular biological systems is often regulated by signaling molecules  called morphogens \cite{stapornwongkul_2021,kicheva_2023}. Morphogens are diffusable chemical substances that are produced in a localized source and can influence cell behaviors or cell fate at a distance in a concentration-dependent manner \cite{lu_2008,nelson_2009}. 
For example, in the development of the mouse mammary gland, a morphogen is produced in the organ and forms concentration gradients away from the gland. Its inhibitory regulation of growth can mediate branching behaviors thereby guiding morphogenesis \cite{sternlicht_2006,daniel_1996,daniel_1989,silberstein_1987,hannezo_2017}. Other biological processes not captured by DLA include degrowth processes and steady-state clusters (homeostasis). Growth, degrowth, and homeostasis play a role in multicellular systems \cite{Palavalli_2021}. Therefore, when developing a lattice model of branching morphogenesis such key features of biological morphogenesis should be taken into account.

In this paper, we present a stochastic lattice model to capture emergent behaviors of branching morphogenesis, including  morphogen-controlled growth and degrowth. Depending on the  length scale of a morphogen gradient, we obtain highly branched DLA-like networks or compact cluster morphologies. We impose topological constraints on cluster dynamics, therefore our networks can grow and degrow while maintaining structural integrity. By combining growth and degrowth, we obtain fluctuating steady-state structures with different scaling properties. Our work presents a minimal model capturing key features of branching morphogenesis that generalizes the classical DLA scenario.

\paragraph{Lattice model of morphogen-controlled growth and degrowth.}---
We represent a multicellular cluster in a surrounding background tissue by occupied sites on a square lattice of size $L\times L$ with lattice spacing $a$. Individual lattice sites have position ${\bf r}$, and neighboring lattice sites are reached moving by a translation vector ${\bf e}$.  An occupation number $n_{\bf r}=1$ ($n_{\bf r}=0$) describes
the presence (absence) of a cluster cell at that site. We denote the morphogen concentration on a lattice site at position ${\bf r}$ by $c_{\bf r}$.

The stochastic dynamics of the occupation numbers is described by the transition rates $\nu^{+}_{\bf r}$ and $\nu^{-}_{\bf r}$ of individual lattice sites.  Growth and degrowth rates are morphogen-dependent and given by
\begin{figure}
    \centering
    \includegraphics{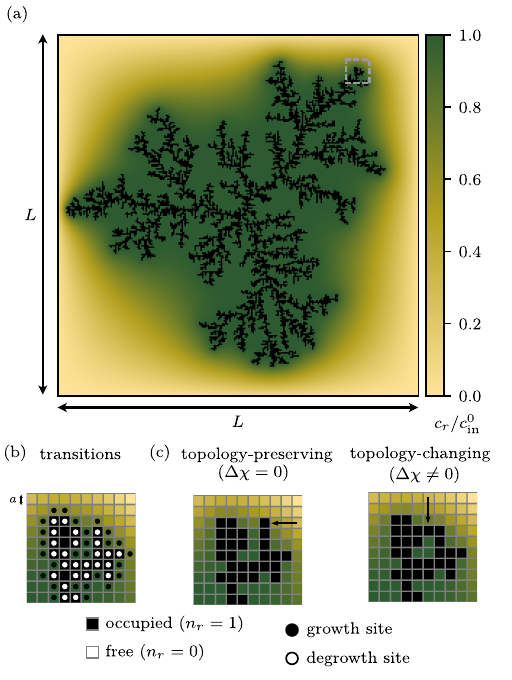}
    \caption{\textbf{Morphogen-controlled cluster growth.} (a) Example cluster of size $N=7000$ with associated morphogen field (green-yellow tone). (b) Magnification of the gray dashed region indicated in (a) with possible growth (filled circle) and degrowth sites (emtpy circle). (c) We present an example of a topology-preserving and topology-changing transition. The changed lattice site is highlighted by an arrow.}
    \label{fig:model_illustration}
\end{figure}

\begin{align}
    \nu^{+}_{\bf r} &= \bar{\nu}^+ \sum_{\bf e} f_{\bf r}(1-n_{\bf r})n_{\bf r+e}\bar{\partial}_{\bf e}^+c_{\bf r}\label{eq:occupation_number_dynamics_growth}\\ 
    \nu^{-}_{\bf r} &= \bar{\nu}^- \sum_{\bf e} f_{\bf r}n_{\bf r} (1-n_{\bf r+e})\bar{\partial}_{\bf e}^-c_{\bf r}\quad .\label{eq:occupation_number_dynamics_degrowth}
\end{align}

These transition rates describe the morphogen-dependent growth an degrowth of the cluster. Empty sites can be filled at rate $\bar{\nu}^+$ and occupied sites can be emptied at rate  ${\nu}^-$. Here, $\bar{\nu}^{\pm}$ are the corresponding space-independent rate coefficients. The operators $\bar{\partial}_{\bf e}^{\pm}c_{\bf r}=\pm(c_{\bf r+e} - c_{\bf r})/a$  denote the lattice gradients of the morphogen, which control growth and degrowth locally. Additionally, the function $f_{\bf r}$ can depend on the occupation number configuration in a local neighborhood around a lattice site. This allows us to introduce additional constraints on  growth or degrowth to conserve the topology of the cluster. 
The morphogen concentration follows a reaction-diffusion dynamics
\begin{equation}
	\partial_t c_{\bf r} = D \bar\Delta c_{\bf r} - k (n_{\bf r})c_{\bf r} + s(n_{\bf r}), \label{eq:morphogen_dynamics}
\end{equation}
where the discrete Laplace operator is $\bar\Delta c_{\bf r} = {a^{-2}}\sum_e[c_{\bf r+e} - c_{\bf r}]$. The morphogen degradation rate $k$ and morphogen production rate $s$ are region-dependent with $k(n_{\bf r}) = k_{\rm in} n_{\bf r} + \kout (1-n_{\bf r})$ and  $s(n_{\bf r}) = \ssin n_{\bf r} + \ssout (1-n_{\bf r})$ where ``in" and ``out" represent inside and outside of the cluster. We study the dynamics of Eq.~\eqref{eq:morphogen_dynamics} subject to the boundary condition $c_{\bm R} = 0$ at the boundaries ${\bm R}$ of the $L\times L$ lattice.

The occupation number dynamics defined by Eq.~\eqref{eq:occupation_number_dynamics_growth}, Eq.~\eqref{eq:occupation_number_dynamics_degrowth} together with the morphogen dynamics Eq.~\eqref{eq:morphogen_dynamics} define a general model for morphogen-controlled cluster growth. Throughout the paper, we consider for simplicity the quasistatic limit where the morphogen field relaxes to its stationary-state faster than growth or degrowth of the cluster. In numerical solutions of our model, we therefore obtain at given time $t$ the stationary-state morphogen field from Eq.~\eqref{eq:morphogen_dynamics} with $\partial_t c_{\bm r}=0$. We then determine the local growth and degrowth rates using Eq.~\eqref{eq:occupation_number_dynamics_growth} and Eq.~\eqref{eq:occupation_number_dynamics_degrowth}. We simulate the growth and degrowth on the lattice using a Kinetic Monte Carlo method, see Appendix~\ref{app:numerical_solution} for details. Fig.~\ref{fig:model_illustration}(a) shows an example  with a cluster and the corresponding morphogen field.

The growth phase of our lattice model with no additional constraints (i.e., $\bar{\nu}^{-}=0$, $f_{\bf r}=1$) can be related to DLA most conveniently in a continuum limit (see Appendix~\ref{app:relation_lattice_model_dla} for a direct correspondence). 
In the continuum limit of DLA, the local growth of clusters results from the statistics of random walkers, which can be described by a diffusion equation with a source on the domain boundary. Similarly, in our model, local growth is related to morphogen gradients, that are determined by reaction-diffusion dynamics. In contrast to DLA, morphogen is produced inside the cluster (as opposed to the domain boundary) and can additionally undergo degradation. 
The diffusion-degradation dynamics introduces length scales $\lambda_{\rm in/out}=\sqrt{D/k_{\rm in/out}}$ that describe the morphogen gradients and leads to a variety of possible  patterns. 
The morphogen concentration in the cluster reaches a value $\cin=s_{\rm in}/k_{\rm in}$ when $\lambdain/a\ll1$. In this case, we identify two extremes with different behavior: 
(i) DLA limit; in the limit $\lambdaout/L\gg1$ of no morphogen degradation and no source outside, $k_{\rm out}=0$,  $s_{\rm out}=0$ and  constant concentration on the cluster $c_{\rm in}$, the dynamics generates a DLA process. (ii) Eden limit; In the limit $\lambdaout/L\ll1$, $s_{\rm out}=0$, the growth is captured by the Eden model, which describes a stochastic growth process with uniform growth rates at the boundary \cite{Eden_1961}. We now extend our model by introducing topological constraints.

\paragraph{Topology-preserving growth and degrowth.}---
During stochastic growth or degrowth, the dynamics can 
introduce changes in cluster topology such as formation of holes or the separation of a cluster in disconnected clusters. The Euler characteristic provides a topological measure that is sensitive to such changes \cite{pratt_2007,yao_2023,flegg_2001}. 
It is defined for planar graphs as $\chi=F - E + V$, where $F$ denotes the number of faces, $E$ denotes the number of edges, and $V$ is the number of vertices.
In space dimension $d=2$ and for the clusters we consider here, the Euler characteristic is equivalent to
\begin{equation}
    \chi=\mathcal{C}-\mathcal{H}\label{eq:euler_characteristic_ch}
\end{equation}
where $\mathcal{C}$ is the number of connected components and $\mathcal{H}$ is the number of holes. Therefore any change in topology is reflected in changes of the Euler characteristic. Given that only a single occupation number $\delta n_{\bf r} = \pm 1$ is changed per event on the lattice, we  obtain the change of Euler characteristic (see Appendix \ref{app:calculation_ec} for a detailed derivation)

\begin{equation}
	\delta \chi = \delta n_{\bf r} \left( 1-\sum_{i=1}^4 m_i\right),\label{eq:change_euler_characteristic}
\end{equation}
where  $ m_i= (1 - n_{e_i})(n_{d_i} + (1 - n_{d_i}) n_{e_{i+1}})$, which takes values $m_i=0$ or $1$. Here, indices $e_i$ and $d_i$ denote nearest and next-nearest neighbors as illustrated in Fig.~\ref{fig:local_neighborhood}.

We can control cluster topology by transition rules
that depend on topology. We introduce topology-preserving transition rates, allowing only transitions that conserve the Euler characteristic. 
 Thus, for $\sum_i m_i=1$ in the local neighborhood of ${\bf r}$, any change at site ${\bf r}$ is topology-preserving since $\delta \chi=0$. We can define a local operator $f(\{m\})$ which is equal to 1 when $\sum_i m_i=1$ and to 0 otherwise. This can be achieved by setting:
\begin{equation}
    f_{\bf r} = \sum_{i=1}^4 m_i\prod_{j\neq i}(1-m_j) \quad .\label{eq:fsq}
\end{equation}
 We next explore our model with numerical examples.

\begin{figure}
    \centering
    \includegraphics{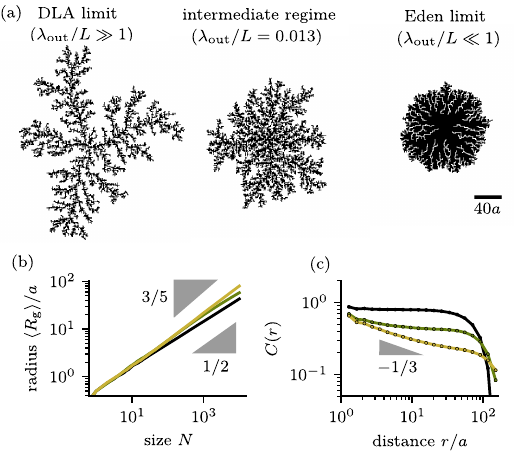}
    \caption{\textbf{Control of cluster morphologies by morphogen dynamics.} (a) Clusters of size $N=10^4$ obtained from Eq.~\eqref{eq:occupation_number_dynamics_growth} and Eq.~\eqref{eq:morphogen_dynamics} for different morphogen gradient lengths $\lambdaout$. (b) Radius of gyration for the structures shown in (a). (c) Correlation function $C(r)$ for the structures shown in (a). The colors correspond to the DLA-limit (yellow), intermediate regime (green), and Eden limit (black). In (b), (c) we averaged over $M=16$ realizations.}  \label{fig:morphology_quantification}
\end{figure}

\paragraph{Tuning growth patterns via morphogens.}---
We first study the growth phase of clusters ($\bar{\nu}^+>0$, $\bar{\nu}^-=0$) using the rule that fixes topology with $f_{\bf r}$ given by Eq.\eqref{eq:fsq}. 
Starting from an initial seed of Euler characteristic $\chi=1$ with $\mathcal{C}=1, \mathcal{H}=0$ (one cluster, no holes), we preserve the topology during the entire time evolution.
We explore the dependence of cluster morphology on the diffusion-degradation length $\lambdaout$. We find a transition from highly branched structures in the DLA limit to compact clusters in the Eden limit (Fig.~\ref{fig:morphology_quantification}a).
In the DLA limit, $\lambda_{\rm out}\gg L$, we find that an instability governs the formation of branched structures, similar to DLA. Morphogen gradients and growth rate are largest at tips. Tip growth increases the morphogen gradient further and a positive feedback sets in leading to the formation of highly branched structures. In the Eden limit with
small diffusion-degradation length,
$\lambda_{\rm out}\ll L$,
 we find a an almost circular morphology with fluctuations. However, in contrast to Eden growth, we observe structures exhibiting lattice-site thin channels that extend deep inside the cluster. Due to the topological constraint of our growth rule, the channels become increasingly unlikely to be filled as they deepen. Filling can occur only through events that take place deep inside the channel, since filling from the exterior of the cluster would create a hole, thereby altering the cluster’s topology.
        
The transition from highly branched to a compact morphology is also reflected in  quantitative measures of cluster morphology. We first study the radius of gyration $R_{\rm g}$ of the cluster 
with $R_{\rm g}^2=N^{-1}\sum_{\bm r} n_{\bm r}\left({\bm r}-\sum_{\bm r} {\bm r}n_{\bm r}/N \right)^2$
as a function of cluster size $N=\sum_{\bm r}n_{\bm r}$. In the DLA and Eden limits, the radius of gyration is well described by power laws 
\begin{equation}
    R_{\rm g} \propto N^{\beta}
\end{equation}
with $\beta=0.57 \pm 0.01$ in the DLA limit and $\beta=0.48\pm 0.01$ in the Eden limit (Fig.~\ref{fig:morphology_quantification}b). This shows that when fixing topology we find scaling behavior that is consistent those obtained without such constraint \cite{meakin_1983,meakin_1983b}. In the intermediate regime, $R_{\rm g}$ crosses over from DLA scaling for small cluster sizes to Eden limit scaling for large cluster sizes. 
This cross-over is governed by the diffusion-degradation length $\lambda_{\rm out}$. We also studied the pairwise correlation function $C(r)=\langle n_{{\bf r}_1} n_{{\bf r}_1+{\bf r}}\rangle_{|{\bf r
}|=r}$ where $\langle \cdot \rangle$ denotes an average over all orientations of ${\bf r}$ and positions ${\bf r}_1$. For $r<R_g$, we find that the  correlation function scales as 
\begin{equation}
    C(r) \propto r^{-\alpha}
\end{equation}
with $\alpha=0.27\pm 0.01$ in the DLA limit and $\alpha=0.043\pm 0.004$ in the Eden limit (Fig.~\ref{fig:morphology_quantification}c), again consistent with previous reports and the relation $\beta=1/(2-\alpha)$ for self-similar systems in two dimensions \cite{witten_1983}.
The correlation function has a sharp decrease for large $r>R_g$ due to the finite size of the cluster.
We also tested whether the scaling exponents are unchanged for variable topology ($f_{\bm r}=1$). We find the same scaling behavior within errors (see Table \ref{tab:exponents}).  

So far we have shown that the morphology of grown clusters can be controlled via diffusion-degradation length $\lambda_{\rm out}$. However, imposing topology has only minor effects in scaling behavior during growth for the DLA limit. We next discuss situations where topology plays a key role during growth and degrowth.

\begin{figure}
    \centering
    \includegraphics{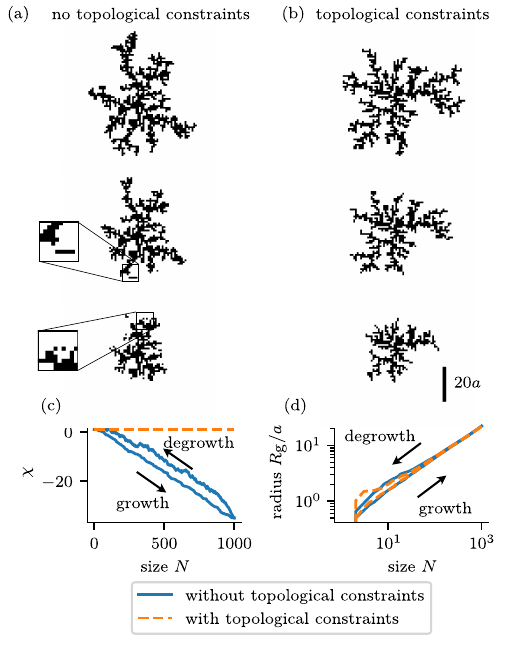}
    \caption{\textbf{Growth and degrowth characterization.} (a) Degrowth of clusters without topological constraints. (b) Degrowth of clusters with topological constraints. (c) Euler characteristic $\chi(t)$ of the structures with and without topological constraint. (d) Radius of gyration $R_{\textrm{g}}$ as a function cluster size $N$ for growth (blue) and degrowth (orange).} 
    \label{fig:growth_degrowth}
\end{figure}

\paragraph{Cluster topology during growth and degrowth.}---
We consider the DLA limit ($k_{\rm out}=0$) to study the distinct effects of introducing the  topological constraint in our model. As a first example, 
we start from a seed in the center of the domain and grow the cluster  until it reaches size $N=100$ ($\bar{\nu}^+>0$, $\bar{\nu}^-=0$). Subsequently we degrow this cluster ($\bar{\nu}^+=0$, $\bar{\nu}^->0$).  We perform this growth-degrowth procedure for two cases: with and without topological constraints.

The growth process without topological constraints, i.e., $f_{\bm r}=1$, leads to the formation of connected components that contains loops (Fig.~\ref{fig:growth_degrowth}a, top). During the degrowth process, loops can be opened and branches can pinch off leading to several disconnected components (Fig.~\ref{fig:growth_degrowth}a, middle and bottom). For the growth-degrowth process with topological constraints using \eqref{eq:fsq}, growth results in a single connected tree-like structure (Fig.~\ref{fig:growth_degrowth}b, top). During degrowth, one connected component is maintained, which is characterized by long, elongated tips as branches are not allowed to pinch off (Fig.~\ref{fig:growth_degrowth}b, middle and bottom). 

We quantify topological and morphological features. The initial condition corresponds to a single connected cluster,  $\chi=1$ as follows from Eq.~\eqref{eq:euler_characteristic_ch}. 
Without topological constraints, the Euler characteristic shrinks during growth, $\chi\leq 1$, since $\mathcal{C}=1$ and $\mathcal{H}\geq1$. During degrowth $\chi$ increases again on a different trajectory (Fig.~\ref{fig:growth_degrowth}c). For fixed topology, we find $\chi=1$ for both growth and degrowth as the topology is maintained to be tree-like. 
For large cluster sizes, the radius of gyration is similar for cases with and without topological constraint. 
However, for small cluster sizes the radius of gyration for fixed topology is larger compared to the case without topological constraints, due to the elongated tips (Fig.~\ref{fig:growth_degrowth}d).
Moreover, we observe that in general the radius of gyration $R_g$ is not only determined by the cluster size $N$. It is different during growth and degrowth phases particularly for smaller $N$. This indicates a history dependence and reveals the irreversibility of the growth and degrowth process.

\begin{figure}
    \centering
    \includegraphics{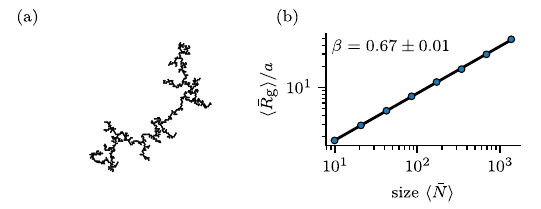}
    \caption{\textbf{Steady-state clusters on a hexagonal lattice} (a) We show a steady-state snapshot of a cluster subject to alternating growth and degrowth on a hexagonal lattice. The shown cluster has size $N=692$. (b) Radius of gyration as a function of stationary system size $N$ averaged over $M=10$ realizations.}
    \label{fig:steady_state_hex}
\end{figure}

\paragraph{Steady-state clusters.}---
\begin{figure}[b]
    \centering
    \includegraphics{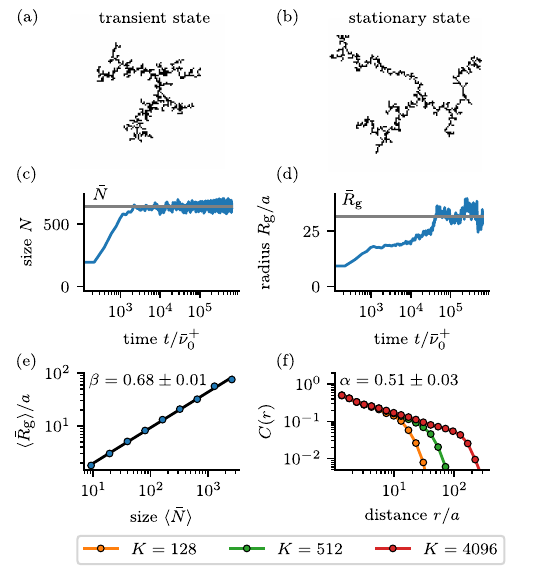}
    \caption{\textbf{Statistical properties of steady-state cluster obtained by alternating growth and degrowth.} (a) We show a snapshot of a cluster subject to alternating growth and degrowth for $t/\bar{\nu}_0^+\approx 3.96\cdot 10^3$. The shown cluster has size $N=599$. (b) We show a snapshot of a cluster subject to alternating growth and degrowth for $t/\bar{\nu}_0^+\approx 1.1\cdot10^5$. The shown cluster has size $N=669$. (c) Cluster size $N(t)$ as a function of time $t$. (d) Radius of gyration $R_{\textrm{g}}$ as a function of time $t$. (e) Average radius of gyration $\langle \bar{R}_{\textrm{g}}\rangle$ as a function of stationary system size $\langle\bar{N}\rangle$. (f) Correlation function $C(r)$ as a function of distance $r$ for the values. The displayed scaling exponent was determined for the case $K=4096$ (red). In (e), (f) we averaged over $M=10$ realizations.}
    \label{fig:steady_state}
\end{figure}
We finally study the scenario of growth and degrowth being present simultaneously ($\bar{\nu}^+>0$, $\bar{\nu}^->0$) in the DLA limit ($k_{\rm out}=0$) with the topological constraints keeping a single connected cluster ($\mathcal{C}=1,\ \mathcal{H}=0$). To study the formation of branched patterns in this scenario, we introduce the logistic growth rate $\bar{\nu}^+=\bar{\nu}_0^+(1-N/K)$ and fix $\bar {\nu}^-$. Here $K$ denotes the carrying capacity and $\bar{\nu}_0^+$ is a parameter (see Table II in Appendix). 
For logistic growth, the cluster first grows until it reaches its steady-state size $\bar{N}$ (Fig.~\ref{fig:steady_state}).
It then evolves from a DLA-like cluster into an elongated structure (Fig.~\ref{fig:steady_state}a,b), while the radius of gyration still increases  until a stationary state is reached where $\bar{R}_{\rm g}$ takes its steady-state value (Fig.~\ref{fig:steady_state}d). To characterize the scaling properties of steady-state clusters, we study the steady-state radius of gyration as a function of the steady-state size (Fig.~\ref{fig:steady_state}e), revealing a power law $\langle \bar{R}_{g}\rangle\sim \langle \bar{N}\rangle^\beta$ with exponent $\beta= 0.68 \pm 0.01$ where the brackets denote averaging over different realizations. Additionally, we determine the correlation function $C$ with the associated scaling exponent $\alpha=0.51\pm0.03$ (for $K=4096$) for steady-state clusters (Fig.~\ref{fig:steady_state}f). 
In both cases, the scaling exponents differ from the scaling exponents obtained for growing structures indicating a different fractal structure. We tested whether the changed exponent is triggered by the anisotropy of the square lattice and studied the same growth-degrowth process on a hexagonal lattice (see Appendix \ref{app:calculation_ec}). An example of a steady-state cluster configuration in a hexagonal lattice is shown in Fig.~\ref{fig:steady_state_hex}(a). In this case, we obtain scaling with exponent $\beta=0.67\pm 0.01$. Thus, the scaling behavior does not change measurably with lattice symmetry. In Table~\ref{tab:exponents} we summarize the scaling exponents obtained in this work together with values reported earlier for related models. 

\begin{table}[]
 \caption{Values of radius of gyration scaling exponent $\beta$ of various systems in $d=2$ in comparison to our results in this work. The growth phase recovers DLA and Eden growth at two extremes controlled by the diffusion-degradation length of morphogen field. The effect of topological rule is negligible in scaling of these two extremes during the growth phase. However, the steady-state clusters with  topological control result in less compact structures than DLA exhibiting a new scaling behavior.}
 \centering
\begin{tabular}{lc}
  \toprule
  & $\beta$ \\
  \midrule
  \multicolumn{2}{l}{\textit{Previous work}} \\[2pt]
  \quad DLA (finite square lattice) \cite{meakin_1983}              & $0.592 \pm 0.017$ \\
  \quad Eden growth (finite square lattice) \cite{meakin_1983b}              & $0.493 \pm 0.002$ \\
  \quad DLA of clusters (finite square lattice) \cite{kolb_1983}             & $0.73  \pm 0.04$  \\
  \quad Branched polymers (theoretical) \cite{isaacson1980flory}    & $0.625$           \\
    \quad Branched polymers (lattice animals) \cite{hsu2005simulations}    & $0.641$           \\
  \quad Self-avoiding random walk (theoretical) \cite{mckenzie1976polymers} & $0.75$ \\
  \midrule
  \multicolumn{2}{l}{\textit{Morphogen-controlled growth (square lattice)}} \\[2pt]
  \quad DLA limit, variable topology   & $0.58 \pm 0.02$ \\
  \quad Eden limit, variable topology  & $0.48 \pm 0.01$ \\
  \quad DLA limit, fixed topology      & $0.57 \pm 0.01$ \\
  \quad Eden limit, fixed topology     & $0.48 \pm 0.01$ \\
  \midrule
  \multicolumn{2}{l}{\textit{Growth–degrowth steady state}} \\[2pt]
  \quad DLA limit (square lattice)     & $0.68 \pm 0.01$ \\
  \quad DLA limit (hexagonal lattice)  & $0.67 \pm 0.01$ \\
  \bottomrule
\end{tabular}
 
\label{tab:exponents}
\end{table}

\paragraph{Conclusion and discussion.}---
In this work, we introduced a stochastic lattice model of branching patterns motivated by biological branching morphogenesis. In this model, pattern formation of cluster emerges from the interplay between the cluster morphology and the  concentration of a morphogen that regulates growth and degrowth.  We impose a topological constraint on cluster morphology during dynamics by introducing a local rule that is sensitive to topological changes. 

The combination of morphogen-controlled growth with topological constraints results in rich pattern formation behaviors and scaling. We show that the range of the morphogen gradient determines whether the cluster becomes highly branched fractal or compact covering DLA limit to Eden limit of stochastic growth processes. A local topological operator allows us to grow and subsequently degrow clusters while conserving the tree-like topology.  Such control over topology can be relevant for  various multicellular systems. On the one hand, it is remarkable that a local operator can maintain global topological characteristic. On the other hand, an interesting question will be to investigate what type of biochemical processes could perform this function.

We have shown that the combined growth-degrowth process leads to steady-state fractal clusters with interesting scaling properties. The exponent $\beta=0.67-0.68$ that we obtain in this case differs from those found during growth and from other random growth/walk processes (see Table \ref{tab:exponents}). Remarkably, it resides at the theoretical ceiling for maximal extension in two-dimensional DLA processes derived by Kesten 
\cite{kesten1987long,kesten1990upper}. This limit has been approached in systems with uniaxial growth along the biased axis \cite{ball1985anisotropy}. A similar trend is observed under lattice anisotropy (square lattice), where the effective exponent of DLA clusters 
shifts from $\beta\approx0.592$
towards $\beta\approx 0.61$
 and has been conjectured to approach 2/3
asymptotically \cite{meakin1986universality,meakin1987structure}; yet even bias-free simulations of up to $10^8$ particles fall short of this asymptotic value \cite{loh2014bias,grebenkov2017anisotropy}. By contrast, our growth-degrowth process reaches $\beta\approx 2/3$ already at $10-3\cdot 10^3$ particles (cells) both for square and hexagonal lattices, suggesting that the bound is saturated not by symmetry breaking but rather because the steady-state topology-preserving dynamics drives the cluster to explore its maximally extended configurations.

Overall, our model bridges concepts from non-equilibrium physics to 
developmental biology and can serve as a starting point to explore generic features of biological fractal branching processes such as vascular or neuronal networks \cite{lorthois_2010, Palavalli_2021}. Given the importance of differential growth of organs during growth \cite{d1917growth}, it will be interesting to include domain growth into our model and study its influence on the formation of patterns \cite{hanauer2026model,bordeu_2023,smith_2019,uccar2023self}.

\begin{acknowledgments}
We would like to thank Jochen Rink, Omar Adame-Arana, and Vincent Hakim for insightful discussions. We acknowledge support from the Max Planck Computing and Data Facility. This work received support from the French
government under the France 2030 investment plan, as part of the Initiative d’Excellence d’Aix-Marseille Université - Amidex (AMX-23-CEI-064).  
\end{acknowledgments}

\bibliography{bibliography}

\appendix
\input{appendix}
\end{document}

%% file: appendix.tex
%\onecolumngrid
\clearpage
%\widetext
% \begin{center}
% \textbf{\large Supplemental Information for ``Control and reversibility of pattern formation in a lattice model of branching morphogenesis''}
% \end{center}
%%%%%%%%%% Merge with supplemental materials %%%%%%%%%%
%%%%%%%%%% Prefix a "S" to all equations, figures, tables and reset the counter %%%%%%%%%%
% \setcounter{equation}{0}
% \setcounter{figure}{0}
% \setcounter{table}{0}
% % \setcounter{page}{1}
% \makeatletter
% \renewcommand{\theequation}{\arabic{equation}}
% \renewcommand{\thefigure}{\arabic{figure}}
% \renewcommand{\bibnumfmt}[1]{[#1]}
% \renewcommand{\citenumfont}[1]{#1}

\section{Numerical solution of the model}
\label{app:numerical_solution}
In our model, we assume morphogen field relaxes faster than the growth-degrowth process, therefore we first update stationary morphogen concentration for a given lattice configuration and then perform a stochastic algorithm. 

To obtain the morphogen concentration at lattice sites $\bf r$, we numerically solve Eq.~\eqref{eq:morphogen_dynamics}. In the limit of quasistatic morphogen dynamics Eq.~\eqref{eq:morphogen_dynamics} reduces to
\begin{equation}
    0 = D \bar{\Delta} c_{\bf r} - k (n_{\bf r})c_{\bf r} + s(n_{\bf r}). \label{eq:morphogen_dynamics_qs}
\end{equation}
This equation defines a set of coupled linear equations that can be rewritten in the form 
\begin{equation}
    0 = A {\bf c} + {\bf s},
\end{equation}
where the vector $\bf c$ contains the morphogen concentrations and the vector $\bf s$ contains the corresponding morphogen source. We further introduced the matrix $A$ that captures the flux of morphogen across lattice sites and also the degradation at lattice sites. $A$ is a positive, semidefinite matrix and therefore we solve this equation using the conjugate gradient method \cite{press_1992,shewchuk_1994}.

To simulate the stochastic growth-degrwoth process defined by the transition rates in Eq.~\eqref{eq:occupation_number_dynamics_growth} and Eq.~\eqref{eq:occupation_number_dynamics_degrowth}, we employ the Gillespie algorithm \cite{gillespie_1977,gillespie_2007}. The Gillespie algorithm separates the problem of obtaining samples of a stochastic process into determining the time at which an event takes places and subsequently determining the type of event that takes place.
To this end, we consider flipping events are non-overlapping. Then, the total rate $\alpha$ for an event to take place is given by
\begin{equation}
	\alpha = \sum_{r} \alpha_{r}  ,
\end{equation}
where $\alpha_{r} = \nu_{r}^+ + \nu_{r}^-$ denotes the rate for an event to take place at lattice site $r$ and the sum runs over all lattice sites ${\bf r}$. Next, we determine the time step $\Delta t$ until the next event from
\begin{equation}
	\Delta t = t - t_0 = -\frac{\log \xi}{\alpha},
\end{equation}
where $\xi$ is a uniformly distributed random number with $\xi\in [0,1]$.
Finally, we determine the lattice site undergoing a transition from
\begin{equation}
	\sum_{{s}=0}^{{r}-1} \alpha_{ s} < \zeta < \sum_{{s}=0}^{{r}} \alpha_{s},
\end{equation}
where $\zeta$ is a uniformly distributed random number with $\zeta\in [0,\alpha]$ and $r$ indicates the chosen lattice site.

Table \ref{table:1} shows the parameters used in this paper.

\section{Correspondence between DLA and morphogen-controlled growth model}
\label{app:relation_lattice_model_dla}

There is a direct mapping of the lattice DLA model to the morphogen-controlled growth model presented in this work in the quasi-stationary limit of the reaction-diffusion process. In the quasi-stationary limit of DLA, the diffusion of particles is a much faster process than the cluster growth, therefore the concentration field of particles $u_{\bf {r} }$ relaxes to steady-state  before the next cluster process is performed. Thus at each time step, 
\begin{equation}
    D\Delta u_{\bf r} =0 \label{eq:dladiff}
\end{equation}
where $D$ is the diffusion coefficient with boundary conditions $u_{ \bf R}=1$ at the system boundary and $u_{\rm in}=0$ inside the cluster. Described as a discrete-time Markov process, the flipping of a site from unoccupied to occupied occurs with a probability proportional to 
\begin{equation}
    {p^{+}_{\bf r}}^{\rm DLA} \propto \sum_{\bf e} (1-n_{\bf r})n_{\bf r+e}u_{\bf r}\quad . \label{eq:dlacluster}
\end{equation}

In a special case of our model, reaction-diffusion parameters $k_{\rm out}=0$, $s/k_{\rm in}=1$, the quasi-stationary $c_{\bf r}$ field reaches 
\begin{equation}
    D\Delta c_{\bf r} =0
\end{equation}
with boundary conditions $c_{\rm in}=1$ inside the cluster and $c_{\bf R}=0$. Now, let us consider a discrete-time version of our model without degrowth ($\bar{\nu}^-=0$). Given a flipping event takes place, the probability that site ${\bf r}$ would be the one flipping is given by 

\begin{equation}
    p^{+}_{\bf r} \propto \sum_{\bf e} (1-n_{\bf r})n_{\bf r+e}(c_{\bf r+e}-c_{\bf r})\quad . \label{eq:ppmorph}
\end{equation}
where we used \eqref{eq:occupation_number_dynamics_growth}. Moreover,  $c_{\bf r+e}=\bar{c}_{\rm in}$ is constant for $n_{\bf r+e}\neq 0$ inside the cluster. Using this in \eqref{eq:ppmorph} and factoring out $\bar{c}_{\rm in}$ suggests 
\begin{equation}
    p^{+}_{\bf r} \propto \sum_{\bf e} (1-n_{\bf r})n_{\bf r+e}(1-c_{\bf r}/\bar{c}_{\rm in})\quad . 
\end{equation}

Thus, the transformation $c_{\bf r}=\bar{c}_{\rm in}(1-u_{\bf r})$ leads exactly to the process described by \eqref{eq:dladiff}-\eqref{eq:dlacluster}.

\section{Calculation of the Euler characteristic}
\label{app:calculation_ec}
For a surface that consists of $F$ faces, $E$ edges, and $V$ vertices the Euler characteristic $\chi$ is defined as
\begin{equation}
    \chi = F - E + V.\label{eq:definition_euler_characteristic_fev}
\end{equation}
Here, we now use this definition to determine the Euler characteristic $\chi$ for the configurations of occupation numbers and the change of the Euler characteristic $\delta \chi$ for the change of a single occupation number. This can be done by calculating changes in number of faces $F$, number of edges $E$, and number of vertices $V$ due to addition or removal of a site at $\bf r$,  $\delta n_{\bf r}=\pm 1$. Below we show the calculations for square and hexagonal lattices.

\begin{figure}[h]
    \includegraphics{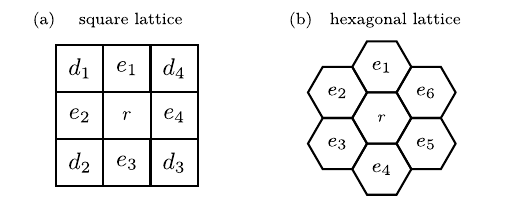}
    \caption{\textbf{Notation for a local neighborhood around a lattice site $\bf r$.} (a) Square lattice: $e_i$ indicate nearest neighbors, while $d_i$ indicate next-nearest neighbors for $i=1\dots4$. (b) Hexagonal lattice: $e_i$ indicate nearest neighbors for $i=1\dots6$.}
    \label{fig:local_neighborhood}
\end{figure}

\textit{In square lattice}, we can write
\begin{eqnarray}
	\delta F_{\bf r} &=& \delta n_{\bf r} , \\  \delta E_{\bf r} &=& \delta n_{\bf r}\sum_{i=1}^4 (1-n_{e_i}), \\  \delta V_{\bf r} &=&  \delta n_{\bf r}\sum_{i=1}^4 (1-n_{e_i}) (1-n_{d_i}) (1-n_{e_{i+1}}),\label{eq:fev_local_definition}
\end{eqnarray}
where $n_{e_i}$ denotes the occupation number of a nearest neighbor and $n_{d_i}$ denotes the occupation number of a next-nearest neighbor (Fig.~\ref{fig:local_neighborhood}(a)).
The first term, $\delta F$, corresponds to the addition/removal of a face by addition/removal of a lattice site. The second term, $\delta E$, reflects the addition/removal of an edge only when the nearest neighbor site is unoccupied. The third one, $\delta V$, shows a vertex is added only when all the three sites sharing a vertex with site ${\bf r}$ are unoccupied. As a result the change in Euler characteristic can be obtained by using \eqref{eq:definition_euler_characteristic_fev}, i.e., $\delta \chi_{\bf r}=\delta F_{\bf r}-\delta E_{\bf r}+\delta V_{\bf r}$ and hence
\begin{eqnarray}
    \delta \chi_{\bf r} &=& \delta n_{\bf r} ( 1 - \sum_{i=1}^4 m_i ), \label{eq:delxir}\\m_i&=& (1 - n_{e_i})(n_{d_i} + (1 - n_{d_i}) n_{e_{i+1}}) \ .\nonumber
\end{eqnarray}

Thus, when $\sum_{i=1}^{4} m_i=1$ in local $3\times3$ neighborhood of ${\bf r}$, the Euler characteristic remains invariant with $\delta n_{\bf r}$, i.e., $\delta \chi_{\bf r}=0$.

\textit{In hexagonal lattice}, the changes in $F$, $E$, and $V$ can be  expressed as:
\begin{eqnarray}
	\delta F_{\bf r} &=& \delta n_{\bf r} , \\  \delta E_{\bf r} &=& \delta n_{\bf r}\sum_{i=1}^6 (1-n_{e_i}), \\  \delta V_{\bf r} &=&  \delta n_{\bf r}\sum_{i=1}^6 (1-n_{e_i}) (1-n_{e_{i+1}}),\label{eq:hexfev_local_definition}
\end{eqnarray}
where $n_{e_i}$ denotes the occupation number of a nearest neighbor (Fig.~
\ref{fig:local_neighborhood}(b)). As a result, \eqref{eq:delxir} is modified for the hexagonal lattice and becomes:

\begin{eqnarray}
    \delta \chi_{\bf r}^{\rm hex}
     &=& \delta n_{\bf r} ( 1 - \sum_{i=1}^6 m_i^{\rm hex} ), \label{eq:hexdelxir}\\m_i^{\rm hex}&=& (1 - n_{e_i}) n_{e_{i+1}}.\nonumber
\end{eqnarray}

The corresponding topological operator is then 

\begin{equation}
    f_{\bf r}^{\rm hex} = \sum_{i=1}^6 m_i^{\rm hex}\prod_{i\neq j}(1-m_j^{\rm hex})  \label{eq:fhex}
\end{equation}
analogous to Eq. \eqref{eq:fsq}.

\newcommand{\paramsfigtwo}{%
  \begin{tabular}[c]{@{}l@{}}
    $3.8\cdot10^{-1}$\\
    $2.4\cdot10^{-3}$\\
    $3.8\cdot10^{-11}$
  \end{tabular}%
}
\newcommand{\paramsfigseven}{%
  \begin{tabular}[c]{@{}l@{}}
    $3.81\cdot 10^{-10}$\\
    $1.53 \cdot 10^{-5}$\\
    $3.81\cdot10^2$
  \end{tabular}%
}

\begin{table*}
	\caption{Parameters used in this paper. The square brackets indicate a range of parameters.}
	\centering
	\begin{tabular}{cc ccc c cc c} \toprule
		 & & \multicolumn{3}{c}{Growth/degrowth simulations} & & & \multicolumn{2}{c}{Steady-state simulations} \\
		 \cmidrule(lr){3-5} \cmidrule(lr){8-9}
		 {Symbol} & {Unit} & {Fig.~\ref{fig:model_illustration}} & {Fig.~\ref{fig:morphology_quantification}} & {Fig.~\ref{fig:growth_degrowth}} & {Symbol} & {Unit} & {Fig.~\ref{fig:steady_state}} & {Fig.~\ref{fig:steady_state_hex}}\\ \midrule
		 $L$ &  $/$ & 250 & 512 & 100 & $L$ & $a$ & 384 & 384  \\
		 $a$ & $a$ & 1 & 1 & 1 & $a$ & $a$ & 1 & 1  \\		
		 $N$ & $/$  & 7000 & $10^4$ & $1,10^3$ & $N$ & / & / & /   \\ \midrule
		 $\bar{\nu}^{+}$   & $\bar{\nu}^{+}$ & 1 & 1 & $1,0$ & $\bar{\nu}_0^{+}$ & $\bar{\nu}^{+}_0$ & 1 & 1 \\
	   $\bar{\nu}^{-}$  & $\bar{\nu}^{+}$ & 0 & 0 & $0,1$ & $\bar{\nu}_{-}$ & $\bar{\nu}^{+}_0$ & 0.5 & 0.5  \\ 
      $K$  & / & /  & / & / & $K$ & / & $[2^4-2^{12}]$ & $[2^4-2^{12}]$  \\ \midrule
		 $D$   & $a^2 \bar{\nu}^{+}$ & 1 & 1 & 1 & $D$ & $a^2 \bar{\nu}^{+}_0$ & 1 & 1/3 \\
		 $\kin$   & $\bar{\nu}^{+}$ & $10^4$ & $10^4$ & $10^5$ & $\kin$ & $\bar{\nu}^{+}_0$ & $10^4$ & $10^4$  \\
		 $\kout$  & $\bar{\nu}^{+}$ & 0  & \paramsfigtwo & 0 & $\kout$ & $\bar{\nu}^{+}_0$ & 0 & 0 \\
		 $\ssin$  & $\bar{\nu}^{+}/a^2$ & $10^4$  & 1 & $10^5$ & $\ssin$ & $\bar{\nu}^{+}_0/a^2$ & $10^4$ & $3\cdot10^4$   \\
		 $\ssout$  & $\bar{\nu}^{+}/a^2$ & 0  & 0 & 0 & $\ssout$ & $\bar{\nu}^{+}_0/a^2$ & 0 & 0  \\ \bottomrule
	\end{tabular}\label{table:1}
\end{table*}

%% file: main_new.bbl
\begin{thebibliography}{69}%
\makeatletter
\providecommand \@ifxundefined [1]{%
 \@ifx{#1\undefined}
}%
\providecommand \@ifnum [1]{%
 \ifnum #1\expandafter \@firstoftwo
 \else \expandafter \@secondoftwo
 \fi
}%
\providecommand \@ifx [1]{%
 \ifx #1\expandafter \@firstoftwo
 \else \expandafter \@secondoftwo
 \fi
}%
\providecommand \natexlab [1]{#1}%
\providecommand \enquote  [1]{``#1''}%
\providecommand \bibnamefont  [1]{#1}%
\providecommand \bibfnamefont [1]{#1}%
\providecommand \citenamefont [1]{#1}%
\providecommand \href@noop [0]{\@secondoftwo}%
\providecommand \href [0]{\begingroup \@sanitize@url \@href}%
\providecommand \@href[1]{\@@startlink{#1}\@@href}%
\providecommand \@@href[1]{\endgroup#1\@@endlink}%
\providecommand \@sanitize@url [0]{\catcode `\\12\catcode `\$12\catcode
  `\&12\catcode `\#12\catcode `\^12\catcode `\_12\catcode `\%12\relax}%
\providecommand \@@startlink[1]{}%
\providecommand \@@endlink[0]{}%
\providecommand \url  [0]{\begingroup\@sanitize@url \@url }%
\providecommand \@url [1]{\endgroup\@href {#1}{\urlprefix }}%
\providecommand \urlprefix  [0]{URL }%
\providecommand \Eprint [0]{\href }%
\providecommand \doibase [0]{https://doi.org/}%
\providecommand \selectlanguage [0]{\@gobble}%
\providecommand \bibinfo  [0]{\@secondoftwo}%
\providecommand \bibfield  [0]{\@secondoftwo}%
\providecommand \translation [1]{[#1]}%
\providecommand \BibitemOpen [0]{}%
\providecommand \bibitemStop [0]{}%
\providecommand \bibitemNoStop [0]{.\EOS\space}%
\providecommand \EOS [0]{\spacefactor3000\relax}%
\providecommand \BibitemShut  [1]{\csname bibitem#1\endcsname}%
\let\auto@bib@innerbib\@empty
%</preamble>
\bibitem [{\citenamefont {Lorthois}\ and\ \citenamefont
  {Cassot}(2010)}]{lorthois_2010}%
  \BibitemOpen
  \bibfield  {author} {\bibinfo {author} {\bibfnamefont {S.}~\bibnamefont
  {Lorthois}}\ and\ \bibinfo {author} {\bibfnamefont {F.}~\bibnamefont
  {Cassot}},\ }\bibfield  {title} {{\selectlanguage {en}\bibinfo {title}
  {Fractal analysis of vascular networks: {Insights} from morphogenesis}},\
  }\href@noop {} {\bibfield  {journal} {\bibinfo  {journal} {Journal of
  Theoretical Biology}\ }\textbf {\bibinfo {volume} {262}},\ \bibinfo {pages}
  {614} (\bibinfo {year} {2010})}\BibitemShut {NoStop}%
\bibitem [{\citenamefont {Family}\ \emph {et~al.}(1989)\citenamefont {Family},
  \citenamefont {Masters},\ and\ \citenamefont {Platt}}]{family_1989}%
  \BibitemOpen
  \bibfield  {author} {\bibinfo {author} {\bibfnamefont {F.}~\bibnamefont
  {Family}}, \bibinfo {author} {\bibfnamefont {B.~R.}\ \bibnamefont
  {Masters}},\ and\ \bibinfo {author} {\bibfnamefont {D.~E.}\ \bibnamefont
  {Platt}},\ }\bibfield  {title} {{\selectlanguage {en}\bibinfo {title}
  {Fractal pattern formation in human retinal vessels}},\ }\href@noop {}
  {\bibfield  {journal} {\bibinfo  {journal} {Physica D: Nonlinear Phenomena}\
  }\textbf {\bibinfo {volume} {38}},\ \bibinfo {pages} {98} (\bibinfo {year}
  {1989})}\BibitemShut {NoStop}%
\bibitem [{\citenamefont {Mainster}(1990)}]{mainster_1990}%
  \BibitemOpen
  \bibfield  {author} {\bibinfo {author} {\bibfnamefont {M.~A.}\ \bibnamefont
  {Mainster}},\ }\bibfield  {title} {{\selectlanguage {en}\bibinfo {title} {The
  {Fractal} {Properties} of {Retinal} {Vessels}: {Embryological} and {Clinical}
  {Implications}}},\ }\href@noop {} {\bibfield  {journal} {\bibinfo  {journal}
  {Eye}\ }\textbf {\bibinfo {volume} {4}},\ \bibinfo {pages} {235} (\bibinfo
  {year} {1990})}\BibitemShut {NoStop}%
\bibitem [{\citenamefont {Weibel}(1963)}]{weibel_1963}%
  \BibitemOpen
  \bibfield  {author} {\bibinfo {author} {\bibfnamefont {E.~R.}\ \bibnamefont
  {Weibel}},\ }\href@noop {} {{\selectlanguage {en}\emph {\bibinfo {title}
  {Morphometry of the {Human} {Lung}}}}}\ (\bibinfo  {publisher} {Springer
  Berlin Heidelberg},\ \bibinfo {address} {Berlin, Heidelberg},\ \bibinfo
  {year} {1963})\BibitemShut {NoStop}%
\bibitem [{\citenamefont {Weibel}(1991)}]{weibel_1991}%
  \BibitemOpen
  \bibfield  {author} {\bibinfo {author} {\bibfnamefont {E.~R.}\ \bibnamefont
  {Weibel}},\ }\bibfield  {title} {{\selectlanguage {en}\bibinfo {title}
  {Fractal geometry: a design principle for living organisms}},\ }\href@noop {}
  {\bibfield  {journal} {\bibinfo  {journal} {American Journal of
  Physiology-Lung Cellular and Molecular Physiology}\ }\textbf {\bibinfo
  {volume} {261}},\ \bibinfo {pages} {L361} (\bibinfo {year}
  {1991})}\BibitemShut {NoStop}%
\bibitem [{\citenamefont {Shree}\ \emph {et~al.}(2022)\citenamefont {Shree},
  \citenamefont {Sutradhar}, \citenamefont {Trottier}, \citenamefont {Tu},
  \citenamefont {Liang},\ and\ \citenamefont {Howard}}]{shree_2022}%
  \BibitemOpen
  \bibfield  {author} {\bibinfo {author} {\bibfnamefont {S.}~\bibnamefont
  {Shree}}, \bibinfo {author} {\bibfnamefont {S.}~\bibnamefont {Sutradhar}},
  \bibinfo {author} {\bibfnamefont {O.}~\bibnamefont {Trottier}}, \bibinfo
  {author} {\bibfnamefont {Y.}~\bibnamefont {Tu}}, \bibinfo {author}
  {\bibfnamefont {X.}~\bibnamefont {Liang}},\ and\ \bibinfo {author}
  {\bibfnamefont {J.}~\bibnamefont {Howard}},\ }\bibfield  {title}
  {{\selectlanguage {en}\bibinfo {title} {Dynamic instability of dendrite tips
  generates the highly branched morphologies of sensory neurons}},\ }\href@noop
  {} {\bibfield  {journal} {\bibinfo  {journal} {Science Advances}\ }\textbf
  {\bibinfo {volume} {8}},\ \bibinfo {pages} {eabn0080} (\bibinfo {year}
  {2022})}\BibitemShut {NoStop}%
\bibitem [{\citenamefont {West}(1997)}]{west_1997}%
  \BibitemOpen
  \bibfield  {author} {\bibinfo {author} {\bibfnamefont {G.~B.}\ \bibnamefont
  {West}},\ }\bibfield  {title} {{\selectlanguage {en}\bibinfo {title} {A
  {General} {Model} for the {Origin} of {Allometric} {Scaling} {Laws} in
  {Biology}}},\ }\href@noop {} {\bibfield  {journal} {\bibinfo  {journal}
  {Science}\ }\textbf {\bibinfo {volume} {276}},\ \bibinfo {pages} {122}
  (\bibinfo {year} {1997})}\BibitemShut {NoStop}%
\bibitem [{\citenamefont {Banavar}\ \emph {et~al.}(2000)\citenamefont
  {Banavar}, \citenamefont {Colaiori}, \citenamefont {Flammini}, \citenamefont
  {Maritan},\ and\ \citenamefont {Rinaldo}}]{banavar_2000}%
  \BibitemOpen
  \bibfield  {author} {\bibinfo {author} {\bibfnamefont {J.~R.}\ \bibnamefont
  {Banavar}}, \bibinfo {author} {\bibfnamefont {F.}~\bibnamefont {Colaiori}},
  \bibinfo {author} {\bibfnamefont {A.}~\bibnamefont {Flammini}}, \bibinfo
  {author} {\bibfnamefont {A.}~\bibnamefont {Maritan}},\ and\ \bibinfo {author}
  {\bibfnamefont {A.}~\bibnamefont {Rinaldo}},\ }\bibfield  {title}
  {{\selectlanguage {en}\bibinfo {title} {Topology of the {Fittest}
  {Transportation} {Network}}},\ }\href@noop {} {\bibfield  {journal} {\bibinfo
   {journal} {Physical Review Letters}\ }\textbf {\bibinfo {volume} {84}},\
  \bibinfo {pages} {4745} (\bibinfo {year} {2000})}\BibitemShut {NoStop}%
\bibitem [{\citenamefont {Katifori}\ \emph {et~al.}(2010)\citenamefont
  {Katifori}, \citenamefont {Szöllősi},\ and\ \citenamefont
  {Magnasco}}]{katifori_2010}%
  \BibitemOpen
  \bibfield  {author} {\bibinfo {author} {\bibfnamefont {E.}~\bibnamefont
  {Katifori}}, \bibinfo {author} {\bibfnamefont {G.~J.}\ \bibnamefont
  {Szöllősi}},\ and\ \bibinfo {author} {\bibfnamefont {M.~O.}\ \bibnamefont
  {Magnasco}},\ }\bibfield  {title} {{\selectlanguage {en}\bibinfo {title}
  {Damage and {Fluctuations} {Induce} {Loops} in {Optimal} {Transport}
  {Networks}}},\ }\href@noop {} {\bibfield  {journal} {\bibinfo  {journal}
  {Physical Review Letters}\ }\textbf {\bibinfo {volume} {104}},\ \bibinfo
  {pages} {048704} (\bibinfo {year} {2010})}\BibitemShut {NoStop}%
\bibitem [{\citenamefont {Corson}(2010)}]{corson_2010}%
  \BibitemOpen
  \bibfield  {author} {\bibinfo {author} {\bibfnamefont {F.}~\bibnamefont
  {Corson}},\ }\bibfield  {title} {{\selectlanguage {en}\bibinfo {title}
  {Fluctuations and {Redundancy} in {Optimal} {Transport} {Networks}}},\
  }\href@noop {} {\bibfield  {journal} {\bibinfo  {journal} {Physical Review
  Letters}\ }\textbf {\bibinfo {volume} {104}},\ \bibinfo {pages} {048703}
  (\bibinfo {year} {2010})}\BibitemShut {NoStop}%
\bibitem [{\citenamefont {Rink}(2013)}]{rink2013stem}%
  \BibitemOpen
  \bibfield  {author} {\bibinfo {author} {\bibfnamefont {J.~C.}\ \bibnamefont
  {Rink}},\ }\bibfield  {title} {\bibinfo {title} {Stem cell systems and
  regeneration in planaria},\ }\href@noop {} {\bibfield  {journal} {\bibinfo
  {journal} {Development genes and evolution}\ }\textbf {\bibinfo {volume}
  {223}},\ \bibinfo {pages} {67} (\bibinfo {year} {2013})}\BibitemShut
  {NoStop}%
\bibitem [{\citenamefont {Le~Verge-Serandour}\ and\ \citenamefont
  {Alim}(2024)}]{le_verge_2024}%
  \BibitemOpen
  \bibfield  {author} {\bibinfo {author} {\bibfnamefont {M.}~\bibnamefont
  {Le~Verge-Serandour}}\ and\ \bibinfo {author} {\bibfnamefont
  {K.}~\bibnamefont {Alim}},\ }\bibfield  {title} {{\selectlanguage
  {en}\bibinfo {title} {\textit{{Physarum} polycephalum} : {Smart} {Network}
  {Adaptation}}},\ }\href@noop {} {\bibfield  {journal} {\bibinfo  {journal}
  {Annual Review of Condensed Matter Physics}\ }\textbf {\bibinfo {volume}
  {15}},\ \bibinfo {pages} {263} (\bibinfo {year} {2024})}\BibitemShut
  {NoStop}%
\bibitem [{\citenamefont {Langer}(1980)}]{langer_1980}%
  \BibitemOpen
  \bibfield  {author} {\bibinfo {author} {\bibfnamefont {J.~S.}\ \bibnamefont
  {Langer}},\ }\bibfield  {title} {{\selectlanguage {en}\bibinfo {title}
  {Instabilities and pattern formation in crystal growth}},\ }\href@noop {}
  {\bibfield  {journal} {\bibinfo  {journal} {Reviews of Modern Physics}\
  }\textbf {\bibinfo {volume} {52}},\ \bibinfo {pages} {1} (\bibinfo {year}
  {1980})}\BibitemShut {NoStop}%
\bibitem [{\citenamefont {Vicsek}(1992)}]{vicsek_1992}%
  \BibitemOpen
  \bibfield  {author} {\bibinfo {author} {\bibfnamefont {T.}~\bibnamefont
  {Vicsek}},\ }\href@noop {} {{\selectlanguage {en}\emph {\bibinfo {title}
  {Fractal {Growth} {Phenomena}}}}},\ \bibinfo {edition} {2nd}\ ed.\ (\bibinfo
  {publisher} {World Scientific Publishing},\ \bibinfo {address} {Singapore},\
  \bibinfo {year} {1992})\BibitemShut {NoStop}%
\bibitem [{\citenamefont {Meakin}(1998)}]{meakin_1998}%
  \BibitemOpen
  \bibfield  {author} {\bibinfo {author} {\bibfnamefont {P.}~\bibnamefont
  {Meakin}},\ }\href@noop {} {{\selectlanguage {eng}\emph {\bibinfo {title}
  {Fractals, scaling and growth far from equilibrium}}}},\ \bibinfo {edition}
  {1st}\ ed.,\ \bibinfo {series} {Cambridge Nonlinear Science Series}\
  No.~\bibinfo {number} {5}\ (\bibinfo  {publisher} {Cambridge University
  Press},\ \bibinfo {address} {Cambridge},\ \bibinfo {year} {1998})\BibitemShut
  {NoStop}%
\bibitem [{\citenamefont {Witten}\ and\ \citenamefont
  {Sander}(1981)}]{witten_1981}%
  \BibitemOpen
  \bibfield  {author} {\bibinfo {author} {\bibfnamefont {T.~A.}\ \bibnamefont
  {Witten}}\ and\ \bibinfo {author} {\bibfnamefont {L.~M.}\ \bibnamefont
  {Sander}},\ }\bibfield  {title} {{\selectlanguage {en}\bibinfo {title}
  {Diffusion-{Limited} {Aggregation}, a {Kinetic} {Critical} {Phenomenon}}},\
  }\href@noop {} {\bibfield  {journal} {\bibinfo  {journal} {Physical Review
  Letters}\ }\textbf {\bibinfo {volume} {47}},\ \bibinfo {pages} {1400}
  (\bibinfo {year} {1981})}\BibitemShut {NoStop}%
\bibitem [{\citenamefont {Witten}\ and\ \citenamefont
  {Sander}(1983)}]{witten_1983}%
  \BibitemOpen
  \bibfield  {author} {\bibinfo {author} {\bibfnamefont {T.~A.}\ \bibnamefont
  {Witten}}\ and\ \bibinfo {author} {\bibfnamefont {L.~M.}\ \bibnamefont
  {Sander}},\ }\bibfield  {title} {{\selectlanguage {en}\bibinfo {title}
  {Diffusion-limited aggregation}},\ }\href@noop {} {\bibfield  {journal}
  {\bibinfo  {journal} {Physical Review B}\ }\textbf {\bibinfo {volume} {27}},\
  \bibinfo {pages} {5686} (\bibinfo {year} {1983})}\BibitemShut {NoStop}%
\bibitem [{\citenamefont {Niemeyer}\ \emph {et~al.}(1984)\citenamefont
  {Niemeyer}, \citenamefont {Pietronero},\ and\ \citenamefont
  {Wiesmann}}]{niemeyer_1984}%
  \BibitemOpen
  \bibfield  {author} {\bibinfo {author} {\bibfnamefont {L.}~\bibnamefont
  {Niemeyer}}, \bibinfo {author} {\bibfnamefont {L.}~\bibnamefont
  {Pietronero}},\ and\ \bibinfo {author} {\bibfnamefont {H.~J.}\ \bibnamefont
  {Wiesmann}},\ }\bibfield  {title} {{\selectlanguage {en}\bibinfo {title}
  {Fractal {Dimension} of {Dielectric} {Breakdown}}},\ }\href@noop {}
  {\bibfield  {journal} {\bibinfo  {journal} {Physical Review Letters}\
  }\textbf {\bibinfo {volume} {52}},\ \bibinfo {pages} {1033} (\bibinfo {year}
  {1984})}\BibitemShut {NoStop}%
\bibitem [{\citenamefont {Paterson}(1984)}]{paterson_1984}%
  \BibitemOpen
  \bibfield  {author} {\bibinfo {author} {\bibfnamefont {L.}~\bibnamefont
  {Paterson}},\ }\bibfield  {title} {{\selectlanguage {en}\bibinfo {title}
  {Diffusion-{Limited} {Aggregation} and {Two}-{Fluid} {Displacements} in
  {Porous} {Media}}},\ }\href@noop {} {\bibfield  {journal} {\bibinfo
  {journal} {Physical Review Letters}\ }\textbf {\bibinfo {volume} {52}},\
  \bibinfo {pages} {1621} (\bibinfo {year} {1984})}\BibitemShut {NoStop}%
\bibitem [{\citenamefont {Måløy}\ \emph {et~al.}(1985)\citenamefont
  {Måløy}, \citenamefont {Feder},\ and\ \citenamefont
  {Jøssang}}]{maloy_1985}%
  \BibitemOpen
  \bibfield  {author} {\bibinfo {author} {\bibfnamefont {K.~J.}\ \bibnamefont
  {Måløy}}, \bibinfo {author} {\bibfnamefont {J.}~\bibnamefont {Feder}},\
  and\ \bibinfo {author} {\bibfnamefont {T.}~\bibnamefont {Jøssang}},\
  }\bibfield  {title} {{\selectlanguage {en}\bibinfo {title} {Viscous
  {Fingering} {Fractals} in {Porous} {Media}}},\ }\href@noop {} {\bibfield
  {journal} {\bibinfo  {journal} {Physical Review Letters}\ }\textbf {\bibinfo
  {volume} {55}},\ \bibinfo {pages} {2688} (\bibinfo {year}
  {1985})}\BibitemShut {NoStop}%
\bibitem [{\citenamefont {Matsushita}\ \emph {et~al.}(1984)\citenamefont
  {Matsushita}, \citenamefont {Sano}, \citenamefont {Hayakawa}, \citenamefont
  {Honjo},\ and\ \citenamefont {Sawada}}]{matsushita1984fractal}%
  \BibitemOpen
  \bibfield  {author} {\bibinfo {author} {\bibfnamefont {M.}~\bibnamefont
  {Matsushita}}, \bibinfo {author} {\bibfnamefont {M.}~\bibnamefont {Sano}},
  \bibinfo {author} {\bibfnamefont {Y.}~\bibnamefont {Hayakawa}}, \bibinfo
  {author} {\bibfnamefont {H.}~\bibnamefont {Honjo}},\ and\ \bibinfo {author}
  {\bibfnamefont {Y.}~\bibnamefont {Sawada}},\ }\bibfield  {title} {\bibinfo
  {title} {Fractal structures of zinc metal leaves grown by
  electrodeposition},\ }\href@noop {} {\bibfield  {journal} {\bibinfo
  {journal} {Phys. Rev. Lett.}\ }\textbf {\bibinfo {volume} {53}},\ \bibinfo
  {pages} {286} (\bibinfo {year} {1984})}\BibitemShut {NoStop}%
\bibitem [{\citenamefont {Fujikawa}\ and\ \citenamefont
  {Matsushita}(1989)}]{fujikawa1989fractal}%
  \BibitemOpen
  \bibfield  {author} {\bibinfo {author} {\bibfnamefont {H.}~\bibnamefont
  {Fujikawa}}\ and\ \bibinfo {author} {\bibfnamefont {M.}~\bibnamefont
  {Matsushita}},\ }\bibfield  {title} {\bibinfo {title} {Fractal growth of
  bacillus subtilis on agar plates},\ }\href@noop {} {\bibfield  {journal}
  {\bibinfo  {journal} {Journal of the physical society of japan}\ }\textbf
  {\bibinfo {volume} {58}},\ \bibinfo {pages} {3875} (\bibinfo {year}
  {1989})}\BibitemShut {NoStop}%
\bibitem [{\citenamefont {Matsushita}\ and\ \citenamefont
  {Fujikawa}(1990)}]{matsushita1990diffusion}%
  \BibitemOpen
  \bibfield  {author} {\bibinfo {author} {\bibfnamefont {M.}~\bibnamefont
  {Matsushita}}\ and\ \bibinfo {author} {\bibfnamefont {H.}~\bibnamefont
  {Fujikawa}},\ }\bibfield  {title} {\bibinfo {title} {Diffusion-limited growth
  in bacterial colony formation},\ }\href@noop {} {\bibfield  {journal}
  {\bibinfo  {journal} {Physica A: Statistical Mechanics and its Applications}\
  }\textbf {\bibinfo {volume} {168}},\ \bibinfo {pages} {498} (\bibinfo {year}
  {1990})}\BibitemShut {NoStop}%
\bibitem [{\citenamefont {Fujikawa}\ and\ \citenamefont
  {Matsushita}(1991)}]{fujikawa1991bacterial}%
  \BibitemOpen
  \bibfield  {author} {\bibinfo {author} {\bibfnamefont {H.}~\bibnamefont
  {Fujikawa}}\ and\ \bibinfo {author} {\bibfnamefont {M.}~\bibnamefont
  {Matsushita}},\ }\bibfield  {title} {\bibinfo {title} {Bacterial fractal
  growth in the concentration field of nutrient},\ }\href@noop {} {\bibfield
  {journal} {\bibinfo  {journal} {Journal of the Physical Society of Japan}\
  }\textbf {\bibinfo {volume} {60}},\ \bibinfo {pages} {88} (\bibinfo {year}
  {1991})}\BibitemShut {NoStop}%
\bibitem [{\citenamefont {Kesten}(1987)}]{kesten1987long}%
  \BibitemOpen
  \bibfield  {author} {\bibinfo {author} {\bibfnamefont {H.}~\bibnamefont
  {Kesten}},\ }\bibfield  {title} {\bibinfo {title} {How long are the arms in
  dla?},\ }\href@noop {} {\bibfield  {journal} {\bibinfo  {journal} {Journal of
  Physics A: Mathematical and General}\ }\textbf {\bibinfo {volume} {20}},\
  \bibinfo {pages} {L29} (\bibinfo {year} {1987})}\BibitemShut {NoStop}%
\bibitem [{\citenamefont {Meakin}(1995)}]{meakin1995progress}%
  \BibitemOpen
  \bibfield  {author} {\bibinfo {author} {\bibfnamefont {P.}~\bibnamefont
  {Meakin}},\ }\bibfield  {title} {\bibinfo {title} {Progress in dla
  research},\ }\href@noop {} {\bibfield  {journal} {\bibinfo  {journal}
  {Physica D: Nonlinear Phenomena}\ }\textbf {\bibinfo {volume} {86}},\
  \bibinfo {pages} {104} (\bibinfo {year} {1995})}\BibitemShut {NoStop}%
\bibitem [{\citenamefont {Halsey}(2000)}]{halsey2000diffusion}%
  \BibitemOpen
  \bibfield  {author} {\bibinfo {author} {\bibfnamefont {T.~C.}\ \bibnamefont
  {Halsey}},\ }\bibfield  {title} {\bibinfo {title} {Diffusion-limited
  aggregation: A model for pattern formation},\ }\href@noop {} {\bibfield
  {journal} {\bibinfo  {journal} {Physics Today}\ }\textbf {\bibinfo {volume}
  {53}},\ \bibinfo {pages} {36} (\bibinfo {year} {2000})}\BibitemShut {NoStop}%
\bibitem [{\citenamefont {Grebenkov}\ and\ \citenamefont
  {Beliaev}(2017)}]{grebenkov2017anisotropy}%
  \BibitemOpen
  \bibfield  {author} {\bibinfo {author} {\bibfnamefont {D.~S.}\ \bibnamefont
  {Grebenkov}}\ and\ \bibinfo {author} {\bibfnamefont {D.}~\bibnamefont
  {Beliaev}},\ }\bibfield  {title} {\bibinfo {title} {How anisotropy beats
  fractality in two-dimensional on-lattice diffusion-limited-aggregation
  growth},\ }\href@noop {} {\bibfield  {journal} {\bibinfo  {journal} {Physical
  Review E}\ }\textbf {\bibinfo {volume} {96}},\ \bibinfo {pages} {042159}
  (\bibinfo {year} {2017})}\BibitemShut {NoStop}%
\bibitem [{\citenamefont {Tolman}\ and\ \citenamefont
  {Meakin}(1989)}]{tolman1989off}%
  \BibitemOpen
  \bibfield  {author} {\bibinfo {author} {\bibfnamefont {S.}~\bibnamefont
  {Tolman}}\ and\ \bibinfo {author} {\bibfnamefont {P.}~\bibnamefont
  {Meakin}},\ }\bibfield  {title} {\bibinfo {title} {Off-lattice and
  hypercubic-lattice models for diffusion-limited aggregation in
  dimensionalities 2--8},\ }\href@noop {} {\bibfield  {journal} {\bibinfo
  {journal} {Physical Review A}\ }\textbf {\bibinfo {volume} {40}},\ \bibinfo
  {pages} {428} (\bibinfo {year} {1989})}\BibitemShut {NoStop}%
\bibitem [{\citenamefont {Hastings}(1997)}]{hastings1997renormalization}%
  \BibitemOpen
  \bibfield  {author} {\bibinfo {author} {\bibfnamefont {M.~B.}\ \bibnamefont
  {Hastings}},\ }\bibfield  {title} {\bibinfo {title} {Renormalization theory
  of stochastic growth},\ }\href@noop {} {\bibfield  {journal} {\bibinfo
  {journal} {Physical Review E}\ }\textbf {\bibinfo {volume} {55}},\ \bibinfo
  {pages} {135} (\bibinfo {year} {1997})}\BibitemShut {NoStop}%
\bibitem [{\citenamefont {Hastings}\ and\ \citenamefont
  {Levitov}(1998)}]{hastings1998laplacian}%
  \BibitemOpen
  \bibfield  {author} {\bibinfo {author} {\bibfnamefont {M.~B.}\ \bibnamefont
  {Hastings}}\ and\ \bibinfo {author} {\bibfnamefont {L.~S.}\ \bibnamefont
  {Levitov}},\ }\bibfield  {title} {\bibinfo {title} {Laplacian growth as
  one-dimensional turbulence},\ }\href@noop {} {\bibfield  {journal} {\bibinfo
  {journal} {Physica D: Nonlinear Phenomena}\ }\textbf {\bibinfo {volume}
  {116}},\ \bibinfo {pages} {244} (\bibinfo {year} {1998})}\BibitemShut
  {NoStop}%
\bibitem [{\citenamefont {Lubensky}\ and\ \citenamefont
  {Isaacson}(1979)}]{lubensky1979statistics}%
  \BibitemOpen
  \bibfield  {author} {\bibinfo {author} {\bibfnamefont {T.}~\bibnamefont
  {Lubensky}}\ and\ \bibinfo {author} {\bibfnamefont {J.}~\bibnamefont
  {Isaacson}},\ }\bibfield  {title} {\bibinfo {title} {Statistics of lattice
  animals and dilute branched polymers},\ }\href@noop {} {\bibfield  {journal}
  {\bibinfo  {journal} {Physical Review A}\ }\textbf {\bibinfo {volume} {20}},\
  \bibinfo {pages} {2130} (\bibinfo {year} {1979})}\BibitemShut {NoStop}%
\bibitem [{\citenamefont {Parisi}\ and\ \citenamefont
  {Sourlas}(1981)}]{parisi1981critical}%
  \BibitemOpen
  \bibfield  {author} {\bibinfo {author} {\bibfnamefont {G.}~\bibnamefont
  {Parisi}}\ and\ \bibinfo {author} {\bibfnamefont {N.}~\bibnamefont
  {Sourlas}},\ }\bibfield  {title} {\bibinfo {title} {Critical behavior of
  branched polymers and the lee-yang edge singularity},\ }\href@noop {}
  {\bibfield  {journal} {\bibinfo  {journal} {Physical Review Letters}\
  }\textbf {\bibinfo {volume} {46}},\ \bibinfo {pages} {871} (\bibinfo {year}
  {1981})}\BibitemShut {NoStop}%
\bibitem [{\citenamefont {You}\ and\ \citenamefont {van
  Rensburg}(1998)}]{you1998critical}%
  \BibitemOpen
  \bibfield  {author} {\bibinfo {author} {\bibfnamefont {S.}~\bibnamefont
  {You}}\ and\ \bibinfo {author} {\bibfnamefont {E.~J.}\ \bibnamefont {van
  Rensburg}},\ }\bibfield  {title} {\bibinfo {title} {Critical exponents and
  universal amplitude ratios in lattice trees},\ }\href@noop {} {\bibfield
  {journal} {\bibinfo  {journal} {Physical Review E}\ }\textbf {\bibinfo
  {volume} {58}},\ \bibinfo {pages} {3971} (\bibinfo {year}
  {1998})}\BibitemShut {NoStop}%
\bibitem [{\citenamefont {Jensen}\ and\ \citenamefont
  {Guttmann}(2000)}]{jensen2000statistics}%
  \BibitemOpen
  \bibfield  {author} {\bibinfo {author} {\bibfnamefont {I.}~\bibnamefont
  {Jensen}}\ and\ \bibinfo {author} {\bibfnamefont {A.~J.}\ \bibnamefont
  {Guttmann}},\ }\bibfield  {title} {\bibinfo {title} {Statistics of lattice
  animals (polyominoes) and polygons},\ }\href@noop {} {\bibfield  {journal}
  {\bibinfo  {journal} {Journal of Physics A: Mathematical and General}\
  }\textbf {\bibinfo {volume} {33}},\ \bibinfo {pages} {L257} (\bibinfo {year}
  {2000})}\BibitemShut {NoStop}%
\bibitem [{\citenamefont {Hsu}\ \emph {et~al.}(2005)\citenamefont {Hsu},
  \citenamefont {Nadler},\ and\ \citenamefont
  {Grassberger}}]{hsu2005simulations}%
  \BibitemOpen
  \bibfield  {author} {\bibinfo {author} {\bibfnamefont {H.-P.}\ \bibnamefont
  {Hsu}}, \bibinfo {author} {\bibfnamefont {W.}~\bibnamefont {Nadler}},\ and\
  \bibinfo {author} {\bibfnamefont {P.}~\bibnamefont {Grassberger}},\
  }\bibfield  {title} {\bibinfo {title} {Simulations of lattice animals and
  trees},\ }\href@noop {} {\bibfield  {journal} {\bibinfo  {journal} {Journal
  of Physics A: Mathematical and General}\ }\textbf {\bibinfo {volume} {38}},\
  \bibinfo {pages} {775} (\bibinfo {year} {2005})}\BibitemShut {NoStop}%
\bibitem [{\citenamefont {Stapornwongkul}\ and\ \citenamefont
  {Vincent}(2021)}]{stapornwongkul_2021}%
  \BibitemOpen
  \bibfield  {author} {\bibinfo {author} {\bibfnamefont {K.~S.}\ \bibnamefont
  {Stapornwongkul}}\ and\ \bibinfo {author} {\bibfnamefont {J.-P.}\
  \bibnamefont {Vincent}},\ }\bibfield  {title} {{\selectlanguage {en}\bibinfo
  {title} {Generation of extracellular morphogen gradients: the case for
  diffusion}},\ }\href@noop {} {\bibfield  {journal} {\bibinfo  {journal}
  {Nature Reviews Genetics}\ }\textbf {\bibinfo {volume} {22}},\ \bibinfo
  {pages} {393} (\bibinfo {year} {2021})}\BibitemShut {NoStop}%
\bibitem [{\citenamefont {Kicheva}\ and\ \citenamefont
  {Briscoe}(2023)}]{kicheva_2023}%
  \BibitemOpen
  \bibfield  {author} {\bibinfo {author} {\bibfnamefont {A.}~\bibnamefont
  {Kicheva}}\ and\ \bibinfo {author} {\bibfnamefont {J.}~\bibnamefont
  {Briscoe}},\ }\bibfield  {title} {{\selectlanguage {en}\bibinfo {title}
  {Control of {Tissue} {Development} by {Morphogens}}},\ }\href@noop {}
  {\bibfield  {journal} {\bibinfo  {journal} {Annual Review of Cell and
  Developmental Biology}\ }\textbf {\bibinfo {volume} {39}},\ \bibinfo {pages}
  {91} (\bibinfo {year} {2023})}\BibitemShut {NoStop}%
\bibitem [{\citenamefont {Lu}\ and\ \citenamefont {Werb}(2008)}]{lu_2008}%
  \BibitemOpen
  \bibfield  {author} {\bibinfo {author} {\bibfnamefont {P.}~\bibnamefont
  {Lu}}\ and\ \bibinfo {author} {\bibfnamefont {Z.}~\bibnamefont {Werb}},\
  }\bibfield  {title} {{\selectlanguage {en}\bibinfo {title} {Patterning
  {Mechanisms} of {Branched} {Organs}}},\ }\href@noop {} {\bibfield  {journal}
  {\bibinfo  {journal} {Science}\ }\textbf {\bibinfo {volume} {322}},\ \bibinfo
  {pages} {1506} (\bibinfo {year} {2008})}\BibitemShut {NoStop}%
\bibitem [{\citenamefont {Nelson}(2009)}]{nelson_2009}%
  \BibitemOpen
  \bibfield  {author} {\bibinfo {author} {\bibfnamefont {C.~M.}\ \bibnamefont
  {Nelson}},\ }\bibfield  {title} {{\selectlanguage {en}\bibinfo {title}
  {Geometric control of tissue morphogenesis}},\ }\href@noop {} {\bibfield
  {journal} {\bibinfo  {journal} {Biochimica et Biophysica Acta - Molecular
  Cell Research}\ }\textbf {\bibinfo {volume} {1793}},\ \bibinfo {pages} {903}
  (\bibinfo {year} {2009})}\BibitemShut {NoStop}%
\bibitem [{\citenamefont {Sternlicht}\ \emph {et~al.}(2006)\citenamefont
  {Sternlicht}, \citenamefont {Kouros-Mehr}, \citenamefont {Lu},\ and\
  \citenamefont {Werb}}]{sternlicht_2006}%
  \BibitemOpen
  \bibfield  {author} {\bibinfo {author} {\bibfnamefont {M.~D.}\ \bibnamefont
  {Sternlicht}}, \bibinfo {author} {\bibfnamefont {H.}~\bibnamefont
  {Kouros-Mehr}}, \bibinfo {author} {\bibfnamefont {P.}~\bibnamefont {Lu}},\
  and\ \bibinfo {author} {\bibfnamefont {Z.}~\bibnamefont {Werb}},\ }\bibfield
  {title} {{\selectlanguage {en}\bibinfo {title} {Hormonal and local control of
  mammary branching morphogenesis}},\ }\href@noop {} {\bibfield  {journal}
  {\bibinfo  {journal} {Differentiation}\ }\textbf {\bibinfo {volume} {74}},\
  \bibinfo {pages} {365} (\bibinfo {year} {2006})}\BibitemShut {NoStop}%
\bibitem [{\citenamefont {Daniel}\ \emph {et~al.}(1996)\citenamefont {Daniel},
  \citenamefont {Robinson},\ and\ \citenamefont {Silberstein}}]{daniel_1996}%
  \BibitemOpen
  \bibfield  {author} {\bibinfo {author} {\bibfnamefont {C.~W.}\ \bibnamefont
  {Daniel}}, \bibinfo {author} {\bibfnamefont {S.}~\bibnamefont {Robinson}},\
  and\ \bibinfo {author} {\bibfnamefont {G.~B.}\ \bibnamefont {Silberstein}},\
  }\bibfield  {title} {{\selectlanguage {en}\bibinfo {title} {The {Role} of
  {TGF}-$\beta$ in {Patterning} and {Growth} of the {Mammary} {Ductal}
  {Tree}}},\ }\href@noop {} {\bibfield  {journal} {\bibinfo  {journal} {Journal
  of Mammary Gland Biology and Neoplasia}\ }\textbf {\bibinfo {volume} {1}},\
  \bibinfo {pages} {331} (\bibinfo {year} {1996})}\BibitemShut {NoStop}%
\bibitem [{\citenamefont {Daniel}\ \emph {et~al.}(1989)\citenamefont {Daniel},
  \citenamefont {Silberstein}, \citenamefont {Van~Horn}, \citenamefont
  {Strickland},\ and\ \citenamefont {Robinson}}]{daniel_1989}%
  \BibitemOpen
  \bibfield  {author} {\bibinfo {author} {\bibfnamefont {C.~W.}\ \bibnamefont
  {Daniel}}, \bibinfo {author} {\bibfnamefont {G.~B.}\ \bibnamefont
  {Silberstein}}, \bibinfo {author} {\bibfnamefont {K.}~\bibnamefont
  {Van~Horn}}, \bibinfo {author} {\bibfnamefont {P.}~\bibnamefont
  {Strickland}},\ and\ \bibinfo {author} {\bibfnamefont {S.}~\bibnamefont
  {Robinson}},\ }\bibfield  {title} {{\selectlanguage {en}\bibinfo {title}
  {{TGF}-$\beta$1-{Induced} {Inhibition} of {Mouse} {Mammary} {Ductal}
  {Growth}: {Developmental} {Specificity} and {Characterization}}},\
  }\href@noop {} {\bibfield  {journal} {\bibinfo  {journal} {Developmental
  Biology}\ }\textbf {\bibinfo {volume} {135}},\ \bibinfo {pages} {20}
  (\bibinfo {year} {1989})}\BibitemShut {NoStop}%
\bibitem [{\citenamefont {Silberstein}\ and\ \citenamefont
  {Daniel}(1987)}]{silberstein_1987}%
  \BibitemOpen
  \bibfield  {author} {\bibinfo {author} {\bibfnamefont {G.~B.}\ \bibnamefont
  {Silberstein}}\ and\ \bibinfo {author} {\bibfnamefont {C.~W.}\ \bibnamefont
  {Daniel}},\ }\bibfield  {title} {{\selectlanguage {en}\bibinfo {title}
  {Reversible {Inhibition} of {Mammary} {Gland} {Growth} by {Transforming}
  {Growth} {Factor}-$\beta$}},\ }\href@noop {} {\bibfield  {journal} {\bibinfo
  {journal} {Science}\ }\textbf {\bibinfo {volume} {237}},\ \bibinfo {pages}
  {291} (\bibinfo {year} {1987})}\BibitemShut {NoStop}%
\bibitem [{\citenamefont {Hannezo}\ \emph {et~al.}(2017)\citenamefont
  {Hannezo}, \citenamefont {Scheele}, \citenamefont {Moad}, \citenamefont
  {Drogo}, \citenamefont {Heer}, \citenamefont {Sampogna}, \citenamefont {van
  Rheenen},\ and\ \citenamefont {Simons}}]{hannezo_2017}%
  \BibitemOpen
  \bibfield  {author} {\bibinfo {author} {\bibfnamefont {E.}~\bibnamefont
  {Hannezo}}, \bibinfo {author} {\bibfnamefont {C.~L.}\ \bibnamefont
  {Scheele}}, \bibinfo {author} {\bibfnamefont {M.}~\bibnamefont {Moad}},
  \bibinfo {author} {\bibfnamefont {N.}~\bibnamefont {Drogo}}, \bibinfo
  {author} {\bibfnamefont {R.}~\bibnamefont {Heer}}, \bibinfo {author}
  {\bibfnamefont {R.~V.}\ \bibnamefont {Sampogna}}, \bibinfo {author}
  {\bibfnamefont {J.}~\bibnamefont {van Rheenen}},\ and\ \bibinfo {author}
  {\bibfnamefont {B.~D.}\ \bibnamefont {Simons}},\ }\bibfield  {title}
  {{\selectlanguage {en}\bibinfo {title} {A {Unifying} {Theory} of {Branching}
  {Morphogenesis}}},\ }\href@noop {} {\bibfield  {journal} {\bibinfo  {journal}
  {Cell}\ }\textbf {\bibinfo {volume} {171}},\ \bibinfo {pages} {242} (\bibinfo
  {year} {2017})}\BibitemShut {NoStop}%
\bibitem [{\citenamefont {Palavalli}\ \emph {et~al.}(2021)\citenamefont
  {Palavalli}, \citenamefont {Tizón-Escamilla}, \citenamefont {Rupprecht},\
  and\ \citenamefont {Lecuit}}]{Palavalli_2021}%
  \BibitemOpen
  \bibfield  {author} {\bibinfo {author} {\bibfnamefont {A.}~\bibnamefont
  {Palavalli}}, \bibinfo {author} {\bibfnamefont {N.}~\bibnamefont
  {Tizón-Escamilla}}, \bibinfo {author} {\bibfnamefont {J.-F.}\ \bibnamefont
  {Rupprecht}},\ and\ \bibinfo {author} {\bibfnamefont {T.}~\bibnamefont
  {Lecuit}},\ }\bibfield  {title} {\bibinfo {title} {Deterministic and
  stochastic rules of branching govern dendrite morphogenesis of sensory
  neurons},\ }\href@noop {} {\bibfield  {journal} {\bibinfo  {journal} {Current
  Biology}\ }\textbf {\bibinfo {volume} {31}},\ \bibinfo {pages} {459}
  (\bibinfo {year} {2021})}\BibitemShut {NoStop}%
\bibitem [{\citenamefont {Eden}(1961)}]{Eden_1961}%
  \BibitemOpen
  \bibfield  {author} {\bibinfo {author} {\bibfnamefont {M.}~\bibnamefont
  {Eden}},\ }\bibfield  {title} {\bibinfo {title} {A two-dimensional growth
  process}\ }(\bibinfo {year} {1961})\BibitemShut {NoStop}%
\bibitem [{\citenamefont {Pratt}(2007)}]{pratt_2007}%
  \BibitemOpen
  \bibfield  {author} {\bibinfo {author} {\bibfnamefont {W.~K.}\ \bibnamefont
  {Pratt}},\ }\href@noop {} {{\selectlanguage {en}\emph {\bibinfo {title}
  {Digital image processing: {PIKS} {Scientific} inside}}}},\ \bibinfo
  {edition} {4th}\ ed.\ (\bibinfo  {publisher} {Wiley-Interscience},\ \bibinfo
  {address} {Hoboken, NJ},\ \bibinfo {year} {2007})\BibitemShut {NoStop}%
\bibitem [{\citenamefont {Yao}\ \emph {et~al.}(2023)\citenamefont {Yao},
  \citenamefont {He}, \citenamefont {Kang}, \citenamefont {Chao},\ and\
  \citenamefont {He}}]{yao_2023}%
  \BibitemOpen
  \bibfield  {author} {\bibinfo {author} {\bibfnamefont {B.}~\bibnamefont
  {Yao}}, \bibinfo {author} {\bibfnamefont {H.}~\bibnamefont {He}}, \bibinfo
  {author} {\bibfnamefont {S.}~\bibnamefont {Kang}}, \bibinfo {author}
  {\bibfnamefont {Y.}~\bibnamefont {Chao}},\ and\ \bibinfo {author}
  {\bibfnamefont {L.}~\bibnamefont {He}},\ }\bibfield  {title}
  {{\selectlanguage {en}\bibinfo {title} {A {Review} for the {Euler} {Number}
  {Computing} {Problem}}},\ }\href@noop {} {\bibfield  {journal} {\bibinfo
  {journal} {Electronics}\ }\textbf {\bibinfo {volume} {12}},\ \bibinfo {pages}
  {4406} (\bibinfo {year} {2023})}\BibitemShut {NoStop}%
\bibitem [{\citenamefont {Flegg}(2001)}]{flegg_2001}%
  \BibitemOpen
  \bibfield  {author} {\bibinfo {author} {\bibfnamefont {G.}~\bibnamefont
  {Flegg}},\ }\href@noop {} {\emph {\bibinfo {title} {From geometry to
  topology}}},\ \bibinfo {edition} {1st}\ ed.\ (\bibinfo  {publisher} {Dover
  Publications},\ \bibinfo {address} {Mineola, NY},\ \bibinfo {year}
  {2001})\BibitemShut {NoStop}%
\bibitem [{\citenamefont {Meakin}(1983{\natexlab{a}})}]{meakin_1983}%
  \BibitemOpen
  \bibfield  {author} {\bibinfo {author} {\bibfnamefont {P.}~\bibnamefont
  {Meakin}},\ }\bibfield  {title} {{\selectlanguage {en}\bibinfo {title}
  {Diffusion-controlled cluster formation in 2—6-dimensional space}},\
  }\href@noop {} {\bibfield  {journal} {\bibinfo  {journal} {Physical Review
  A}\ }\textbf {\bibinfo {volume} {27}},\ \bibinfo {pages} {1495} (\bibinfo
  {year} {1983}{\natexlab{a}})}\BibitemShut {NoStop}%
\bibitem [{\citenamefont {Meakin}(1983{\natexlab{b}})}]{meakin_1983b}%
  \BibitemOpen
  \bibfield  {author} {\bibinfo {author} {\bibfnamefont {P.}~\bibnamefont
  {Meakin}},\ }\bibfield  {title} {{\selectlanguage {en}\bibinfo {title} {The
  {Vold}-{Sutherland} and {Eden} models of cluster formation}},\ }\href@noop {}
  {\bibfield  {journal} {\bibinfo  {journal} {Journal of Colloid and Interface
  Science}\ }\textbf {\bibinfo {volume} {96}},\ \bibinfo {pages} {415}
  (\bibinfo {year} {1983}{\natexlab{b}})}\BibitemShut {NoStop}%
\bibitem [{\citenamefont {Kolb}\ \emph {et~al.}(1983)\citenamefont {Kolb},
  \citenamefont {Botet},\ and\ \citenamefont {Jullien}}]{kolb_1983}%
  \BibitemOpen
  \bibfield  {author} {\bibinfo {author} {\bibfnamefont {M.}~\bibnamefont
  {Kolb}}, \bibinfo {author} {\bibfnamefont {R.}~\bibnamefont {Botet}},\ and\
  \bibinfo {author} {\bibfnamefont {R.}~\bibnamefont {Jullien}},\ }\bibfield
  {title} {\bibinfo {title} {Scaling of kinetically growing clusters},\
  }\href@noop {} {\bibfield  {journal} {\bibinfo  {journal} {Physical Review
  Letters}\ }\textbf {\bibinfo {volume} {51}},\ \bibinfo {pages} {1123}
  (\bibinfo {year} {1983})}\BibitemShut {NoStop}%
\bibitem [{\citenamefont {Isaacson}\ and\ \citenamefont
  {Lubensky}(1980)}]{isaacson1980flory}%
  \BibitemOpen
  \bibfield  {author} {\bibinfo {author} {\bibfnamefont {J.}~\bibnamefont
  {Isaacson}}\ and\ \bibinfo {author} {\bibfnamefont {T.}~\bibnamefont
  {Lubensky}},\ }\bibfield  {title} {\bibinfo {title} {Flory exponents for
  generalized polymer problems},\ }\href@noop {} {\bibfield  {journal}
  {\bibinfo  {journal} {Journal de Physique Lettres}\ }\textbf {\bibinfo
  {volume} {41}},\ \bibinfo {pages} {469} (\bibinfo {year} {1980})}\BibitemShut
  {NoStop}%
\bibitem [{\citenamefont {McKenzie}(1976)}]{mckenzie1976polymers}%
  \BibitemOpen
  \bibfield  {author} {\bibinfo {author} {\bibfnamefont {D.~S.}\ \bibnamefont
  {McKenzie}},\ }\bibfield  {title} {\bibinfo {title} {Polymers and scaling},\
  }\href@noop {} {\bibfield  {journal} {\bibinfo  {journal} {Physics Reports}\
  }\textbf {\bibinfo {volume} {27}},\ \bibinfo {pages} {35} (\bibinfo {year}
  {1976})}\BibitemShut {NoStop}%
\bibitem [{\citenamefont {Kesten}(1990)}]{kesten1990upper}%
  \BibitemOpen
  \bibfield  {author} {\bibinfo {author} {\bibfnamefont {H.}~\bibnamefont
  {Kesten}},\ }\bibfield  {title} {\bibinfo {title} {Upper bounds for the
  growth rate of dla},\ }\href@noop {} {\bibfield  {journal} {\bibinfo
  {journal} {Physica A: Statistical Mechanics and its Applications}\ }\textbf
  {\bibinfo {volume} {168}},\ \bibinfo {pages} {529} (\bibinfo {year}
  {1990})}\BibitemShut {NoStop}%
\bibitem [{\citenamefont {Ball}\ \emph {et~al.}(1985)\citenamefont {Ball},
  \citenamefont {Brady}, \citenamefont {Rossi},\ and\ \citenamefont
  {Thompson}}]{ball1985anisotropy}%
  \BibitemOpen
  \bibfield  {author} {\bibinfo {author} {\bibfnamefont {R.~C.}\ \bibnamefont
  {Ball}}, \bibinfo {author} {\bibfnamefont {R.~M.}\ \bibnamefont {Brady}},
  \bibinfo {author} {\bibfnamefont {G.}~\bibnamefont {Rossi}},\ and\ \bibinfo
  {author} {\bibfnamefont {B.~R.}\ \bibnamefont {Thompson}},\ }\bibfield
  {title} {\bibinfo {title} {Anisotropy and cluster growth by diffusion-limited
  aggregation},\ }\href@noop {} {\bibfield  {journal} {\bibinfo  {journal}
  {Physical review letters}\ }\textbf {\bibinfo {volume} {55}},\ \bibinfo
  {pages} {1406} (\bibinfo {year} {1985})}\BibitemShut {NoStop}%
\bibitem [{\citenamefont {Meakin}(1986)}]{meakin1986universality}%
  \BibitemOpen
  \bibfield  {author} {\bibinfo {author} {\bibfnamefont {P.}~\bibnamefont
  {Meakin}},\ }\bibfield  {title} {\bibinfo {title} {Universality,
  nonuniversality, and the effects of anisotropy on diffusion-limited
  aggregation},\ }\href@noop {} {\bibfield  {journal} {\bibinfo  {journal}
  {Physical Review A}\ }\textbf {\bibinfo {volume} {33}},\ \bibinfo {pages}
  {3371} (\bibinfo {year} {1986})}\BibitemShut {NoStop}%
\bibitem [{\citenamefont {Meakin}\ \emph {et~al.}(1987)\citenamefont {Meakin},
  \citenamefont {Ball}, \citenamefont {Ramanlal},\ and\ \citenamefont
  {Sander}}]{meakin1987structure}%
  \BibitemOpen
  \bibfield  {author} {\bibinfo {author} {\bibfnamefont {P.}~\bibnamefont
  {Meakin}}, \bibinfo {author} {\bibfnamefont {R.~C.}\ \bibnamefont {Ball}},
  \bibinfo {author} {\bibfnamefont {P.}~\bibnamefont {Ramanlal}},\ and\
  \bibinfo {author} {\bibfnamefont {L.~M.}\ \bibnamefont {Sander}},\ }\bibfield
   {title} {\bibinfo {title} {Structure of large two-dimensional square-lattice
  diffusion-limited aggregates: Approach to asymptotic behavior},\ }\href@noop
  {} {\bibfield  {journal} {\bibinfo  {journal} {Physical Review A}\ }\textbf
  {\bibinfo {volume} {35}},\ \bibinfo {pages} {5233} (\bibinfo {year}
  {1987})}\BibitemShut {NoStop}%
\bibitem [{\citenamefont {Loh}(2014)}]{loh2014bias}%
  \BibitemOpen
  \bibfield  {author} {\bibinfo {author} {\bibfnamefont {Y.~L.}\ \bibnamefont
  {Loh}},\ }\bibfield  {title} {\bibinfo {title} {Bias-free simulation of
  diffusion-limited aggregation on a square lattice},\ }\href@noop {}
  {\bibfield  {journal} {\bibinfo  {journal} {arXiv preprint arXiv:1407.2586}\
  } (\bibinfo {year} {2014})}\BibitemShut {NoStop}%
\bibitem [{\citenamefont {d'Arcy}(1917)}]{d1917growth}%
  \BibitemOpen
  \bibfield  {author} {\bibinfo {author} {\bibfnamefont {T.}~\bibnamefont
  {d'Arcy}},\ }\href@noop {} {\emph {\bibinfo {title} {On growth and form}}}\
  (\bibinfo  {publisher} {Cambridge university press},\ \bibinfo {year}
  {1917})\BibitemShut {NoStop}%
\bibitem [{\citenamefont {Hanauer}\ \emph {et~al.}(2026)\citenamefont
  {Hanauer}, \citenamefont {Palavalli}, \citenamefont {An}, \citenamefont
  {Ilker}, \citenamefont {Rink},\ and\ \citenamefont
  {J{\"u}licher}}]{hanauer2026model}%
  \BibitemOpen
  \bibfield  {author} {\bibinfo {author} {\bibfnamefont {C.}~\bibnamefont
  {Hanauer}}, \bibinfo {author} {\bibfnamefont {A.}~\bibnamefont {Palavalli}},
  \bibinfo {author} {\bibfnamefont {B.}~\bibnamefont {An}}, \bibinfo {author}
  {\bibfnamefont {E.}~\bibnamefont {Ilker}}, \bibinfo {author} {\bibfnamefont
  {J.~C.}\ \bibnamefont {Rink}},\ and\ \bibinfo {author} {\bibfnamefont
  {F.}~\bibnamefont {J{\"u}licher}},\ }\bibfield  {title} {\bibinfo {title}
  {Model for self-organized growth, branching, and allometric scaling of the
  planarian gut},\ }\href@noop {} {\bibfield  {journal} {\bibinfo  {journal}
  {PRX Life}\ }\textbf {\bibinfo {volume} {4}},\ \bibinfo {pages} {013014}
  (\bibinfo {year} {2026})}\BibitemShut {NoStop}%
\bibitem [{\citenamefont {Bordeu}\ \emph {et~al.}(2023)\citenamefont {Bordeu},
  \citenamefont {Chatzeli},\ and\ \citenamefont {Simons}}]{bordeu_2023}%
  \BibitemOpen
  \bibfield  {author} {\bibinfo {author} {\bibfnamefont {I.}~\bibnamefont
  {Bordeu}}, \bibinfo {author} {\bibfnamefont {L.}~\bibnamefont {Chatzeli}},\
  and\ \bibinfo {author} {\bibfnamefont {B.~D.}\ \bibnamefont {Simons}},\
  }\bibfield  {title} {{\selectlanguage {en}\bibinfo {title} {Inflationary
  theory of branching morphogenesis in the mouse salivary gland}},\ }\href@noop
  {} {\bibfield  {journal} {\bibinfo  {journal} {Nature Communications}\
  }\textbf {\bibinfo {volume} {14}},\ \bibinfo {pages} {3422} (\bibinfo {year}
  {2023})}\BibitemShut {NoStop}%
\bibitem [{\citenamefont {Smith}\ \emph {et~al.}(2019)\citenamefont {Smith},
  \citenamefont {Mailler},\ and\ \citenamefont {Yates}}]{smith_2019}%
  \BibitemOpen
  \bibfield  {author} {\bibinfo {author} {\bibfnamefont {C.~A.}\ \bibnamefont
  {Smith}}, \bibinfo {author} {\bibfnamefont {C.}~\bibnamefont {Mailler}},\
  and\ \bibinfo {author} {\bibfnamefont {C.~A.}\ \bibnamefont {Yates}},\
  }\bibfield  {title} {{\selectlanguage {en}\bibinfo {title} {Unbiased
  on-lattice domain growth}},\ }\href@noop {} {\bibfield  {journal} {\bibinfo
  {journal} {Physical Review E}\ }\textbf {\bibinfo {volume} {100}},\ \bibinfo
  {pages} {063307} (\bibinfo {year} {2019})}\BibitemShut {NoStop}%
\bibitem [{\citenamefont {U{\c{c}}ar}\ \emph {et~al.}(2023)\citenamefont
  {U{\c{c}}ar}, \citenamefont {Hannezo}, \citenamefont {Tiilikainen},
  \citenamefont {Liaqat}, \citenamefont {Jakobsson}, \citenamefont {Nurmi},\
  and\ \citenamefont {Vaahtomeri}}]{uccar2023self}%
  \BibitemOpen
  \bibfield  {author} {\bibinfo {author} {\bibfnamefont {M.~C.}\ \bibnamefont
  {U{\c{c}}ar}}, \bibinfo {author} {\bibfnamefont {E.}~\bibnamefont {Hannezo}},
  \bibinfo {author} {\bibfnamefont {E.}~\bibnamefont {Tiilikainen}}, \bibinfo
  {author} {\bibfnamefont {I.}~\bibnamefont {Liaqat}}, \bibinfo {author}
  {\bibfnamefont {E.}~\bibnamefont {Jakobsson}}, \bibinfo {author}
  {\bibfnamefont {H.}~\bibnamefont {Nurmi}},\ and\ \bibinfo {author}
  {\bibfnamefont {K.}~\bibnamefont {Vaahtomeri}},\ }\bibfield  {title}
  {\bibinfo {title} {Self-organized and directed branching results in optimal
  coverage in developing dermal lymphatic networks},\ }\href@noop {} {\bibfield
   {journal} {\bibinfo  {journal} {Nature Communications}\ }\textbf {\bibinfo
  {volume} {14}},\ \bibinfo {pages} {5878} (\bibinfo {year}
  {2023})}\BibitemShut {NoStop}%
\bibitem [{\citenamefont {Press}\ \emph {et~al.}(1992)\citenamefont {Press},
  \citenamefont {Teukolsky}, \citenamefont {Vetterling},\ and\ \citenamefont
  {Flannery}}]{press_1992}%
  \BibitemOpen
  \bibfield  {author} {\bibinfo {author} {\bibfnamefont {W.~H.}\ \bibnamefont
  {Press}}, \bibinfo {author} {\bibfnamefont {S.~A.}\ \bibnamefont
  {Teukolsky}}, \bibinfo {author} {\bibfnamefont {W.~T.}\ \bibnamefont
  {Vetterling}},\ and\ \bibinfo {author} {\bibfnamefont {B.~P.}\ \bibnamefont
  {Flannery}},\ }\href@noop {} {{\selectlanguage {en}\emph {\bibinfo {title}
  {Numerical {Recipes} in {C}: {The} {Art} of {Scientific} {Computing}}}}},\
  \bibinfo {edition} {2nd}\ ed.\ (\bibinfo  {publisher} {Cambridge University
  Press},\ \bibinfo {address} {Cambridge, United Kingdom},\ \bibinfo {year}
  {1992})\BibitemShut {NoStop}%
\bibitem [{\citenamefont {Shewchuk}(1994)}]{shewchuk_1994}%
  \BibitemOpen
  \bibfield  {author} {\bibinfo {author} {\bibfnamefont {J.~R.}\ \bibnamefont
  {Shewchuk}},\ }\href@noop {} {\emph {\bibinfo {title} {An {Introduction} to
  the {Conjugate} {Gradient} {Method} {Without} the {Agonizing} {Pain}}}},\
  \bibinfo {type} {Tech. Rep.}\ (\bibinfo  {institution} {Carnegie-Mellon
  University, Department of Computer Science},\ \bibinfo {address} {Pittsburgh,
  PA},\ \bibinfo {year} {1994})\BibitemShut {NoStop}%
\bibitem [{\citenamefont {Gillespie}(1977)}]{gillespie_1977}%
  \BibitemOpen
  \bibfield  {author} {\bibinfo {author} {\bibfnamefont {D.~T.}\ \bibnamefont
  {Gillespie}},\ }\bibfield  {title} {{\selectlanguage {en}\bibinfo {title}
  {Exact {Stochastic} {Simulation} of {Coupled} {Chemical} {Reactions}}},\
  }\href@noop {} {\bibfield  {journal} {\bibinfo  {journal} {The Journal of
  Physical Chemistry}\ }\textbf {\bibinfo {volume} {81}},\ \bibinfo {pages}
  {2340} (\bibinfo {year} {1977})}\BibitemShut {NoStop}%
\bibitem [{\citenamefont {Gillespie}(2007)}]{gillespie_2007}%
  \BibitemOpen
  \bibfield  {author} {\bibinfo {author} {\bibfnamefont {D.~T.}\ \bibnamefont
  {Gillespie}},\ }\bibfield  {title} {{\selectlanguage {en}\bibinfo {title}
  {Stochastic {Simulation} of {Chemical} {Kinetics}}},\ }\href@noop {}
  {\bibfield  {journal} {\bibinfo  {journal} {Annual Review of Physical
  Chemistry}\ }\textbf {\bibinfo {volume} {58}},\ \bibinfo {pages} {35}
  (\bibinfo {year} {2007})}\BibitemShut {NoStop}%
\end{thebibliography}%
